\RequirePackage[2020-02-02]{latexrelease}
\documentclass[aps,nofootinbib,12pt]{revtex4}
\usepackage{graphicx,color}
\usepackage{relsize}
\usepackage{appendix}
\usepackage{latexsym,amsmath,amssymb,graphicx,booktabs}
\usepackage{hyperref}
\usepackage{cleveref}
\usepackage{caption}

\usepackage{slashed}
\usepackage[makeroom]{cancel}
\usepackage{bbold}
\usepackage{color,soul}
\usepackage%[toc,page]
{appendix}
\numberwithin{equation}{section}
\graphicspath{ {./} }
\begin{document}

\title{\Large\bf Majorana propagator on de Sitter space}
%\date{\today}

\author{\bf  Tomislav Prokopec~\footnote{{\tt email:} t.prokopec@uu.nl}}

\affiliation{Institute for Theoretical Physics and EMME$\Phi$, 
Faculty of Science, Utrecht University, Princetonplein 5, 3584 CC Utrecht, The Netherlands}

\author{\bf Vishnu Hari Unnithan~\footnote{{\tt email:} vishnuhari5@outlook.com, %v.h.unnithan@students.uu.nl
}}
\affiliation{Institute for Theoretical Physics and EMME$\Phi$, Faculty of Science, Utrecht University, Princetonplein 5, 3584 CC Utrecht, The Netherlands}

\begin{abstract}

\noindent
We study the dynamics of Majorana fermions in an expanding de Sitter space and  
to that aim we construct the vacuum Feynman propagator for Majorana fermions 
in de Sitter space. Surprisingly, the Majorana propagator is identical to that of the Dirac fermions. 
We then use this propagator and calculate the one-loop effective action for Majorana fermions, and show that it differs from that of Dirac fermions by a factor of $1/2$,
which accounts for the reduced number of degrees of freedom of the Majorana particle. 
Finally, we derive the Majorana and Dirac propagators for mixing fermionic flavors,
which are suitable for studying CP-violating effects in quantum loops.

\end{abstract}

\pacs{}
\maketitle

%%%%%%%%%%%%%%%%%%%%%%
%%%        I N T R O D U C T I O N         %%%
%%%%%%%%%%%%%%%%%%%%%%

\section{Introduction}
\label{Introduction}

Understanding Majorana particles has attracted considerable interest in the last few decades. Understanding its dynamics have consequences for important questions such as the nature of dark matter~\cite{Chikashige:1980ui,Lattanzi:2014mia,Beltran:2008xg}, the origin and nature of matter anti-matter asymmetry, the origin of neutrino masses, inflationary models, {\it etc}. The field of quantum computing, topological insulators and other fields in low energy physics have also wrestled with Majorana particles~\cite{Beenakker:2011np}, as pairs of Majorana particles could provide topologically protected quantum gates. 

The Dirac fermion propagator on de Sitter space was originally constructed by 
Candelas and Raine~\cite{Candelas:1975du}, and then subsequently generalized to accelerating, homogeneous cosmological spaces in Ref.~\cite{Koksma:2009tc}.~\footnote{The Dirac propagator was constructed in spacetimes of a constant acceleration parameter, $\epsilon=-\dot H/H^2\; (\dot H = {\rm d}H/{\rm d}t)$, where $H(t)$ denotes the expansion rate of the Universe and assuming that the fermionic mass squared scales as the Ricci curvature scalar.}
How to construct Majorana propagators in de Sitter space and in cosmological spaces was discussed in
Refs.~\cite{Cotaescu:2018afn,Cotaescu:2018zrq}, where  the Majorana propagator was obtained by acting the left-handed chiral projectors on the Dirac propagator.

 In this work we provide an alternative derivation of the Majorana propagator by using 
 the method of mode sums and discuss in detail how to impose the Majorana condition. We point out in particular at the difference that arises when Majorana condition is imposed {\it on-shell} and {\it off-shell}, and stress the importance of imposing it off-shell, as it can be important for accurate perturbative calculations. One of the advantages of using the technique of mode sums is that it brings to fore how the reduced number of degree of freedom of Majorana particles influences construction of the propagator, which is not evident in the procedure 
in Refs.~\cite{Cotaescu:2018afn,Cotaescu:2018zrq}. 

Due to the Majorana nature of neutrinos, 
leptogenesis~\cite{Fukugita:1986hr} (see Ref.~\cite{Bodeker:2020ghk} for a review) is a key motivator for understanding the dynamics of Majorana fermions as we know that leptogenesis is concerned with lepton number generation in extensions of the standard model that include heavy Majorana neutrinos. Lepton asymmetry produced {\it via} the neutrinos is important for our understanding of the baryon asymmetry of the Universe, as the one-loop decay processes are studied by making use of the resummed thermal propagator for Majorana neutrinos~\cite{Pilaftsis:2003gt,Pilaftsis:2009pk,Buchmuller:1997yu,Rangarajan:1999kt}. This motivates a careful consideration of the Majorana propagator in out-of-equilibrium conditions in which 
CP-violating processes play an important role, one notable example being time dependent backgrounds of the expanding Universe. Since leptogenesis process involves mixing of lepton flavors, here we also consider the multiflavor Majorana propagator in de Sitter space and compare it with that of the Dirac fermions.

Motivated by the advances in precision observations of cosmic microwave background 
radiation and Universe's large scale structure, in recent decades we have witnessed progress 
in perturbative results 
for interacting matter and gravitational fields both in de Sitter space (which serves 
as the model space for cosmological inflation), 
as well as in more general cosmological spacetimes. The most important ingredient of these studies are the propagators, as they are the essential building block for perturbative studies. Scalar propagators in de Sitter space were first derived in~\cite{Chernikov:1968zm}, also 
see~\cite{Allen:1987tz}, \cite{Fukuma:2013mx}. Next, the vector propagators 
on de Sitter space were initially constructed in covariant Fermi 
gauges in~\cite{Allen:1985wd}, 
for massive vectors of scalar electrodynamics in exact Lorenz gauge in~\cite{Tsamis:2006gj},
and some subtleties related gauge fixing were subsequently discussed in Ref.~\cite{Frob:2013qsa}. 
The graviton propagator was constructed and discussed at length in 
Refs.~\cite{Antoniadis:1986sb,Tsamis:1992xa,Allen:1986ta,Allen:1986tt,Higuchi:2001uv,Miao:2011fc,Mora:2012zi,Frob:2016hkx,Glavan:2019msf}. 
Our knowledge on the propagators in more general cosmological spacetimes is rather limited. 
The massless scalar propagator for inflarionary spaces with constant deceleration parameter 
was obtained in Ref.~\cite{Janssen:2008px}
and the massive vector propagator for scalar electrodynamics was recently 
constructed in Ref.~\cite{Glavan:2020zne}, where the authors discussed the subtleties of 
unitary gauge and various simpler limits of the propagator.
These propagators have been used to study various perturbative processes in inflation,
but the list of references is large, and so we refrain from presenting it here.

The paper is organised as follows. In Section~\ref{Section-2} we define the model that 
we intend to study, and define Majorana fields in the helicity basis and solve for the mode 
functions. Section~\ref{Section-3} is devoted to the construction of the propagator, which we check by showing that it reduces to the correct Minkowski space propagator. In the same section we also investigate how the propagator transforms under charge conjugation, as that is of interest for understanding CP violation. 
As a simple application of the propagator in Section~\ref{Section-4} we compute the one-loop 
effective action and show that it equals 1/2 of that for the Dirac fermions. In order to better understand fermionic CP violation 
(which is important for leptogenesis), in Section~\ref{Section-5} we construct the propagator 
for multiflavor Majorana fermions and compare with the corresponding propagator for Dirac fermions. Our main results are discussed in Section~\ref{Conclusion and Discussion}, where we also present an outlook to possible applications of the propagator. Finally, 
in Section~\ref{Appendices} we present various technical derivations.

%%%%%%%%%%%%%%%%%%%%
%%%           T H E   M O D E L         %%%
%%%%%%%%%%%%%%%%%%%%

\section{The Model}
\label{Section-2}

To describe the dynamics of Majorana fermions in de Sitter space, we begin
by recalling the Lagrangian of Dirac fermions in general gravitational backgrounds in D space-time dimensions,
as this allows one to perform perturbative calculations with 
renormalization and regularisation that respects all the symmetries of the problem. The action and Lagrangian of the free massive Dirac fermion in D dimensions are,
\begin{equation}
    \label{2.1}
    S[\Psi,\bar\Psi] = \int d^Dx\sqrt{-g} {\cal L}
    \;;\quad
   \mathcal{L} = \frac12i\bar{\Psi}\gamma^{\mu}
\!\stackrel{\leftrightarrow}{\nabla}_{\mu}\!\!\Psi - \bar{\Psi}\mathcal{M}\Psi
\,,
\end{equation}
where 
$\Psi(x)$ denotes the Dirac spinor, $\bar\Psi = \Psi^\dagger \gamma^0$,
$\bar{\Psi}\gamma^{\mu}\!\!\stackrel{\leftrightarrow}{\nabla}_{\mu}\!\!\Psi
=\bar{\Psi}\gamma^{\mu}\big[\nabla_{\mu}\Psi\big]
 -\big[\nabla_{\mu}\bar{\Psi}\big]\gamma^{\mu}\Psi$ 
 (we use chiral representation for the Dirac
 matrices, whose basic properies are listed in Appendix~A),
 and $\mathcal{M}$ is the (complex) mass matrix defined as:
\begin{equation}
    \label{2.2}
    \mathcal{M} = (m_{1} \mathbb{1}_{2^{D/2}\times 2^{D/2}} 
                 + im_{2}\gamma_{5}) = \begin{pmatrix}
    m^{*}\mathbb{1}_{2^{(D-2)/2}\times2^{(D-2)/2}}&0\\
    0&m\mathbb{1}_{2^{(D-2)/2}\times2^{(D-2)/2}}
    \end{pmatrix}\,,
\end{equation}
with $m=m_1+im_2$, with $m_1,m_2\in \mathbb R$.
The metric of a spatially flat, homogeneous universe can be written in terms of the 
time-dependent scale factor $a=a(t)$ as, 
\begin{equation}
    \label{2.3}
    g_{\mu\nu} = \text{diag}\left(-1,a^{2},a^{2}....,a^{2}\right)_{\text{D}}
    \,,
\end{equation}
such that its determinant equals, $g=-a^{2(D-1)}$.
The covariant derivative operator $\nabla_{\!\mu}$ acts on a spinor as,
\begin{equation}
    \label{2.4}
    \nabla_{\!\mu}\Psi = \left(\partial_{\mu} - \Gamma_{\mu}\right)\Psi,
\end{equation}
and spin(or) connection $\Gamma_{\mu}$ can be obtained from, 
\begin{equation}
    \label{2.5}
    \Gamma_{\mu} = -\frac{1}{8}e^{\nu}_{c}\left(\partial_{\mu}e_{\nu d} - \Gamma^{\alpha}_{\mu\nu}e_{\alpha d}\right)\left[\gamma^{c},\gamma^{d}\right]\,.
\end{equation}
where $e^{\nu}_{c}$ denote {\it vielbein} (used to project 
vectors (and tensors) on tangent space, $V^\mu(x) = e^\mu_a(x)V^a$, where we use Latin letters to denote indices on flat tangent space).
By transforming to conformal time $\eta$ as,
\begin{equation}
\label{2.6}
    {\rm d}t = a(\eta){\rm d}\eta,
\end{equation}
in which the metric~\label{2.3} becomes, $g_{\mu\nu} = a^2(\eta)\eta_{\mu\nu}$
($\eta_{\mu\nu}$ denotes Minkowski metric in $D$ dimensions) 
and using {\it vielbein} formalism~\cite{Koksma:2009tc} one arrives at,
\begin{equation}
    \label{2.7}
    i\gamma^{\mu}\nabla_{\mu}\Psi(x) = a^{-\frac{D+1}{2}}(\eta)i\gamma^{b}\partial_{b}\left(a^{\frac{D-1}{2}}(\eta)\Psi(x)\right)
\,.
\end{equation}
Here $a(\eta)$ is the scale factor for a spatially flat homogeneous expanding universe and $\eta$ is conformal time. The relation~\eqref{2.7} is very useful as it relates the covariant derivative acting on a spinor to the partial derivative 
in (flat) tangent space.

\subsection{Dirac equation}
\label{Dirac equation}

The Dirac equation which follows from Eq. \eqref{2.1} is given by,
\begin{equation}
    \label{2.8}
    \left(i\gamma^{\mu}\nabla_{\mu} - \mathcal{M}\right)\Psi(x) = 0
    \,.
\end{equation}
Using Eq.~\eqref{2.7} and upon performing a conformal rescaling 
of the fermion field,
\begin{equation}
    \label{2.9}
    \Psi(x)\rightarrow a^{\frac{D-1}{2}}(\eta)\Psi(x) = \Tilde{\Psi}(x)
\,,
\end{equation}
the Dirac equation~\eqref{2.8} can be recast as,
\begin{equation}
    \label{2.10}
    \left(i\gamma^{b}\partial_{b} - a{\mathcal{M}}\right)\Tilde{\Psi}(x) = 0
\,.
\end{equation}

\subsection{Majorana fermions}
\label{Majorana fermions}

The motivation to carefully investigate the dynamics of Majorana particles in de Sitter space also arises, as stated above, from the observation that there is a lack of conserved vector currents for Majorana particles. To see this we can define the Dirac field in terms of its two chiral spinors as follows,
\begin{equation}
    \label{2.11}
    \Psi(x) = \begin{pmatrix}
    \chi_{_{L}}(x)\\
    \chi_{_{R}}(x)
    \end{pmatrix}
\,.
\end{equation}
The Lagrangian~\eqref{2.1} then becomes ({\it cf.} Appendix~A), 
\begin{equation}
    \label{2.12}
    \mathcal{L} = i\chi^{\dagger}_{_{L}}\bar{\sigma}^{\mu}\nabla_{\mu}\chi_{_{L}} 
    + i\chi^{\dagger}_{_{R}}\sigma^{\mu}\nabla_{\mu}\chi_{_{R}} 
    - m^{*}\chi^{\dagger}_{_{R}}\chi_{_{L}} - m\chi^{\dagger}_{_{L}}\chi_{_{R}}
\,,
\end{equation}
where ${\sigma}^{\mu}=(\mathbb{1},\sigma^i)$ and $\bar{\sigma}^{\mu}=(\mathbb{1},-\sigma^i)$,
where $\sigma^i$ denote  Pauli matrices,
$i\sigma^\mu\nabla_\mu\chi_{_{R}} 
=a^{-\frac{D+1}{2}}i{\sigma}^{\mu}\partial_\mu\big[a^{\frac{D-1}{2}}\chi_{_{R}}\big]$
and $i\bar\sigma^\mu\nabla_\mu\chi_{_{L}} 
=a^{-\frac{D+1}{2}}i\bar{\sigma}^{\mu}\partial_\mu\big[a^{\frac{D-1}{2}}\chi_{_{L}}\big]$.
Under a global U(1) transformation, $\chi_{_{L,R}}\rightarrow e^{-i\theta}\chi_{_{L,R}}$, and therefore
the fields $\Psi$  and $\bar\Psi$ transform infinitesimally as, $\Psi\rightarrow \Psi+ \theta \Phi_\theta$
and $\bar\Psi\rightarrow \bar\Psi+ \theta \bar\Phi_\theta$, with $\Phi_\theta = -i \Psi$
and $\bar\Phi_\theta = i \bar\Psi$, from which it follows that the Lagrangian~(\ref{2.12}) remains invariant. 
These transformations imply the following conserved (Noether) vector current for Dirac particles,
\begin{equation}
    \label{2.13}
    j^{\mu}(x) \equiv \frac{\partial {\cal L}}{\partial\partial_\mu \Psi}\Phi_\theta
    + \frac{\partial {\cal L}}{\partial\partial_\mu \bar\Psi}\bar\Phi_\theta
                  = \bar{\Psi}(x)\gamma^{\mu}\Psi(x)
\,.
\end{equation}

Let us now consider the Majorana particles, defined by the Majorana condition:
\begin{equation}
    \label{2.14}
    \Psi(x) = -i\gamma^{2}\Psi^{*}(x) = \begin{pmatrix}
    0 & -\epsilon\\
    \epsilon & 0
    \end{pmatrix}\begin{pmatrix}
    \chi^{*}_{_{L}}(x)\\
    \chi^{*}_{_{R}}(x)
    \end{pmatrix}
    \,.
\end{equation}
Here we have defined $i\sigma^{2} = \epsilon$, where $\epsilon^{ab}$ denotes the antisymmetric tensor in two 
dimensions, defined by $\epsilon^{12}= 1 = -\epsilon^{21}$ and $\epsilon^{11} = \epsilon^{22}=1$.
 Thus, the Majorana field $\Psi_{M}(x)$ is defined as
\begin{equation}
    \label{2.15}
    \Psi_{M}(x) = \begin{pmatrix}
    \chi_{_{L}}(x)\\
    \epsilon\chi^{*}_{_{L}}(x)
    \end{pmatrix}.
\end{equation}
The simplest way to arrive at the Lagrangian for Majorana fermions is to 
to replace $\Psi$ by $\Psi_M$ into~(\ref{2.15}), which results in the following Lagrangian,~\footnote{As argued below, there is a better way to write the Majorana lagrangian as the normalization 
of the kinetic terms in~(\ref{2.16}) is not canonical.}
\begin{equation}
    \label{2.16}
    \mathcal{L}_M = -i\chi_{_{L}}^{T}\epsilon\sigma^{\mu}\epsilon\, \nabla_{\!\mu}\chi^{*}_{_{L}}
    + i\chi^{\dagger}_{_{L}}\bar{\sigma}^{\mu} \nabla_{\!\mu}\chi_{_{L}} 
    + m^{*}\chi^{T}_{_{L}}\epsilon\chi_{_{L}} - m\chi^{\dagger}_{_{L}}\epsilon\chi^{*}_{_{L}},
\end{equation}
It is clear that this Lagrangian is not invariant under the global U(1) transformation discussed above, 
as the Majorana condition~(\ref{2.14})
is inconsistent with the symmetry transformation discussed above.

 An important consequence of this observation is 
that there is {\it no conserved vector current for Majorana fermions.}
%~\footnote{Na\^ ively inserting $\Psi_M$ from 
%Eq.~(\ref{2.15}) into the vector current~(\ref{2.13}) would yield, $j^\mu = 0$, indicating that the 
%vector current for Majorana particles has no physical meaning.}
This can be understood in terms of Clifford algebra and the accompanying symmetry generators for Dirac and Majorana particles. It can be noted that (in $D=4$) the Dirac particles have 16 generators of Clifford algebra whereas the Majorana particles have 15 generators, precisely due to the fact that the Majorana condition~(\ref{2.14})
is inconsistent with the symmetry transformation discussed above. This means that the symmetry algebra of Majorana
particles is the Clifford algebra minus the unity matrix, resulting in the symmetry generators of {\it conformal group}
in $D=4$~\cite{BarrosoMancha:2019}. 
 As a step towards understading ramifications of the lack of conserved vector current for the dynamics 
 of Majorana fermions, in this paper we construct the Majorana propagator in de Sitter space, thereby paying  special attention to the implementation of the Majorana condition~\eqref{2.14}. 
 This will allow us to compare the Majorana propagator with that of Dirac fermions originally constructed by 
Candelas and Raine in Ref.~\cite{Candelas:1975du} and to the more general propagator 
discussed in Ref.~\cite{Koksma:2009tc}. We already know that the absence of vector current plays an important 
role for the CP-violating dynamics of mixing fermions, as the baryogenesis mechanism from Ref.~\cite{Garbrecht:2005rr}
would be fundamentally changed in the absence of a conserved vector current 
({\it cf.} Eq.~(5) of Ref.~\cite{Garbrecht:2005rr}).

\subsubsection{Canonical quantization}

The above considerations and Eq.~(\ref{2.1}) suggest the following 
action and Lagrangian for the Majorana fermions in its canonical form, 
\begin{equation}
    \label{2.1M}
    S_M[\Psi] = \int d^Dx\sqrt{-g} {\cal L}_M
    \;;\quad
    \mathcal{L}_M = \frac14\bar{\Psi}_Mi\gamma^{\mu}
\!\!\stackrel{\leftrightarrow}{\nabla}_{\mu}\!\!\Psi_M - \frac12\bar{\Psi}_M\mathcal{M}\Psi_M
\,,
\end{equation}
The additional factor 1/2 can be justified by noting that -- due to the Majorana condition --
$\Psi_M$ and $\bar\Psi_M$ are not independent fields, which can be clearly seen from, 
\begin{equation}
    \label{2.1M2}
    \Psi_M = (-i\gamma^2)\Psi_M^* =(-i\gamma^2)\gamma^0\left[\bar\Psi_M\right]^T
\, \;\Longrightarrow \; \bar\Psi_M= \Psi_M^T\gamma^0(i\gamma^2) 
\,.
\end{equation}
Since the condition is imposed at the level of the action, we refer to it as 
the {\it off-shell Majorana condition}.
This then means that, varying the action~(\ref{2.1M})
 with respect to $\Psi(x)$ results in the Dirac equation that has the same form as in Eq.~(\ref{2.8}),
 but with the Majorana condition imposed on $\Psi_M$,
\begin{equation}
   \label{2.1M3}
    \left(i\gamma^{\mu}\nabla_{\mu} - \mathcal{M}\right)\Psi_M(x) = 0
    \,,\qquad  \bar\Psi_M= \Psi_M^T\gamma^0(i\gamma^2)
    \,.
\end{equation}
It is now quite easy to show that, transposing~(\ref{2.1M3}) and inserting 
$\gamma^0(i\gamma^2)(-i\gamma^2) \gamma^0 = 1$ and commuting $(-i\gamma^2) \gamma^0$
 with $\big[i(\gamma^{\mu})^T\nabla_{\mu} - \mathcal{M}\big]$ (${\cal M}^T={\cal M}$) results in 
\begin{equation}
   \label{2.1M4}
  \bar \Psi_M(x)\left(\!-i\gamma^{\mu}\!\stackrel{\leftarrow}{\nabla}_{\!\mu} - \mathcal{M}\right) = 0
    \,,
\end{equation}
which is therefore {\it not} independent 
from Eq.~(\ref{2.1M3}).~\footnote{
\label{complex and real scalar}
A simple and well studied analogy
to the situation at hand is the case of complex and real scalar fields. The canonically 
normalized Lagrangian for a complex scalar $\Phi$ is, 
${\cal L}_\Phi=-g^{\mu\nu}(\partial_\mu\Phi^*)(\partial_\nu\Phi)-m^2\Phi^*\Phi$,
where $\Phi^*$ and $\Phi$ are considered independent degrees of freedom.
On the other hand, the canonically normalized Lagrangian for a real scalar field $\phi$ is,
${\cal L}_\phi = -\frac12g^{\mu\nu}(\partial_\mu\phi)(\partial_\nu\phi)-\frac12m^2\phi^2$,
where the reality condition $\phi^*(x)=\phi(x)$ plays the role of the Majorana condition for fermions,
which is just the {\it charge} transformation for scalar fields.}

The canonically normalized Lagrangian for the Majorana field $\Psi_{M}(x)$ in~\eqref{2.1M} 
(when recast in terms of $2^{(D-2)/2}$-spinor fields) reads,
\begin{equation}
    \label{2.1M5}
    \mathcal{L}_M = -\frac{1}{4}\chi_{_{L}}^{T}\epsilon i\sigma^{\mu}\epsilon\,
               \stackrel{\leftrightarrow}{\nabla}_{\!\mu}\chi^{*}_{_{L}} 
    +\frac{1}{4} \chi^{\dagger}_{_{L}}i\bar{\sigma}^{\mu} \stackrel{\leftrightarrow}{\nabla}_{\!\mu}\chi_{_{L}} 
    +\frac{1}{2} m^{*}\chi^{T}_{_{L}}\epsilon\chi_{_{L}}
     -\frac{1}{2} m\chi^{\dagger}_{_{L}}\epsilon\chi^{*}_{_{L}}
\,.
\end{equation}
Notice the 1/2 difference when compared with the na\^ive Majorana Lagrangian~(\ref{2.16}).
The Majorana condition was already used when writing~(\ref{2.1M4})  (to remove $\chi_{_{R}}$),
and therefore
the spinors $\chi_{_{L}}$ and $\chi^{*}_{_{L}}$ in~(\ref{2.1M5}) are independent complex spinor fields. The corresponding canonical momenta are 
obtained by varying the action $S_M$ with respect to $\partial_0\chi_{_{L}}(x)$ and 
$\partial_0\chi_{_{L}}^*(x)$ resulting in 
$\left[\pi_{\chi_L^*}\right]_b = \frac12a^{D-1}i\left[\chi_L^T\right]_b$ and 
$\left[\pi_{\chi_L}\right]_b = \frac12a^{D-1}i\big[\chi_L^\dagger\big]_b$,
where $b$ denotes a spinor index. When these are promoted to operators,
 $\hat{\chi}_{_{L}}(x)$, $\hat{\chi}^{*}_{_{L}}(x)$, $\hat\pi_{\chi_L}$ 
 and $\hat\pi_{\chi_L^*}$,
canonical quantization then implies the following anticommutators ($\hbar =1$), 
\begin{equation}
    \label{2.1M6}
    \begin{split}
       \{\hat{\chi}_{_{L}}(\eta,\vec{x}), \hat\pi_{{\chi}_{_{L}}}(\eta,\vec{x}^{\,\prime})\}_{bc} 
                 &=  i\delta^{D-1}(\vec{x}-\vec{x}^{\,\prime})\delta_{bc} \\
       \{ \hat{\chi}^{*}_{_{L}}(\eta,\vec{x}), \hat\pi_{{\chi}^*_{_{L}}}(\eta,\vec{x}^{\,\prime})\}_{bc} 
             &= i\delta^{D-1}(\vec{x}-\vec{x}^{\,\prime})\delta_{bc},
    \end{split}
\end{equation}
or equivalently, 
\begin{equation}
    \label{2.17}
    \begin{split}
       \{\hat{\chi}_{_{L}}(\eta,\vec{x}), \hat{\chi}^{\dagger}_{_{L}}(\eta,\vec{x}^{\,\prime})\}_{bc} 
                 &= \frac{2\delta^{D-1}(\vec{x}-\vec{x}^{\,\prime})}{a(\eta)^{D-1}}\delta_{bc} \\
       \{ \hat{\chi}^{*}_{_{L}}(\eta,\vec{x}), \hat{\chi}^{T}_{_{L}}(\eta,\vec{x}^{\,\prime})\}_{bc} 
             &=  \frac{2\delta^{D-1}(\vec{x}-\vec{x}^{\,\prime})}{a(\eta)^{D-1}}\delta_{bc},
    \end{split}
\end{equation}
and all other anti-commutators vanish. Note the conspicuous factor 2 on the right hand side
of these relations, which is absent in canonical quantization of Dirac fermions.
This factor is a consequence of our requirement to canonically normalize the kinetic term
for Majorana fermions, and it will affect the normalization of our mode functions, and thus the 
propagator. There is no deep physical meaning in it as this factor can be absorbed by a suitable 
rescaling of $\chi_L$ and $\chi_L^*$. Finally, note that even though the Lagrangian~(\ref{2.1M5}) suggests that there are two independent canonical momenta, the canonical quantization reveals 
that there is only one independent anti-commutator (they are related by transposition). This is to be contrasted with the Dirac fermions, for which there are two independent 
canonical momenta, each of them associated with the left and right handed fermion 
fields, respectively. The Majorana condition imposes a dependency between 
the right- and left-handed fermions, 
thus reducing the number of independent canonical momenta to just one.

\subsubsection{Helicity decomposition of Majorana fields}
\label{Helicity decomposition of Majorana fields}

In this section we represent Majorana spinors in terms of mode sums, where the modes 
are helicity eignespinors. Some attention is given to how to construct helicity eigenspinors in $D$ dimensions, in which they are $2^{(D-2)/2}-$component vectors. 

Using the Majorana condition in Eq.~\eqref{2.14} we can define the (conformally rescaled) Majorana fields as follows,
\begin{eqnarray}
    \label{2.18}
    \hat{\Tilde{\Psi}}_{M,\alpha}(x) 
    \!\!&=&\!\! \sum_{h=\pm}\int \frac{d^{D-1}k}{(2\pi)^{D-1}}
     e^{i\Vec{k}\cdot\Vec{x}}\left[\hat{b}_{h}(\vec{k}) 
        + \hat{b}^{\dagger}_{h}(-\vec{k})\right]A_{h,\alpha}(\eta,\Vec{k}),
\\
 \label{2.18b}
  \hat{\Tilde{\bar\Psi}}_{M,\alpha}(x)
  \!\!&=&\!\! 
  { \sum_{h=\pm}\int \frac{d^{D-1}k}{(2\pi)^{D-1}}
     e^{-i\Vec{k}\cdot\Vec{x}}\left[\hat{b}_{h}(-\vec{k}) 
        + \hat{b}^{\dagger}_{h}(\vec{k})\right]\bar A_{h,\alpha}(\eta,\Vec{k})}
\, , 
\end{eqnarray}
the derivation of which is presented in Appendix~B. $\hat{\Tilde{\Psi}}_{M,\alpha}(x)$
 is the rescaled Majorana field,
\begin{equation}
    \label{2.19}
    \hat{\Tilde{\Psi}}_{M,\alpha}(x) = \begin{pmatrix}
    a^{\frac{D-1}{2}}\hat{\chi}_{_{L}}(x)\\
    \epsilon a^{\frac{D-1}{2}}\hat{\chi}^{*}_{_{L}}(x)
    \end{pmatrix} = \begin{pmatrix}
    \hat{\rho}(x)\\
    \epsilon \hat{\rho}^{*}(x)
    \end{pmatrix}
\,,
\end{equation}
and $\hat{\Tilde{\bar\Psi}}_{M,\alpha}(x)=\hat{\Tilde{\Psi}}^\dagger_{M,\beta}(x)\gamma^0_{\beta\alpha}$.
From~(\ref{2.17}) one infers the (only) nontrivial anti-commutation relation for the rescaled spinors,
\begin{equation}
    \label{2.20}
    \begin{split}
       \{\hat{\rho}(\eta,\vec{x}),\hat{\rho}^{\dagger}(\eta,\vec{x}^{\,\prime})\}_{ab} = \{\hat{\chi}_{_{L}}(\eta,\vec{x}), a^{D-1}\hat{\chi}^{\dagger}_{_{L}}(\eta,\vec{x}^{\,\prime})\}_{ab} 
        &= 2\delta^{D-1}(\vec{x}-\vec{x}^{\,\prime})\delta_{ab}\\
 %      \{\hat{\rho}^{*}(\eta,\vec{x}),\hat{\rho}^{T}(\eta,\vec{x}^{\,\prime})\}_{ab} = \{ \hat{\chi}^{*}_{_{L}}(\eta,\vec{x}), a^{D-1}\hat{\chi}^{T}_{_{L}}(\eta,\vec{x}^{\,\prime})\}_{ab} 
 %         &= \delta^{D-1}(\vec{x}-\vec{x}^{\,\prime})\delta_{ab},
    \end{split}
\end{equation}
The operators $\hat{b}_{h}(\vec{k})$ and $\hat{b}^\dagger_{h}(\vec{k})$ in the mode decomposition~(\ref{2.18})
 are the Majorana fermion annihilation and creation operators, respectively. $\hat{b}_{h}(\vec{k})$
 annihilates the vacuum state
$\lvert \Omega\rangle$,
 $\hat{b}_{h}(\vec{k})\lvert \Omega\rangle = 0$, and $\hat{b}^\dagger_{h}(\vec{k})$
creates a particle of momentum $\vec k$ and helicity $h$, and can be used to construct 
states that span the Hilbert space of the problem. Note that the creation and annihilation 
operators in Eq.~(\ref{2.18}) are identical, which is due to the fact that the positive and negative 
frequency states (related by charge conjugation) do not independently fluctuate, as it is dictated by 
the Majorana condition.
 The operators obey the following anti-commutation relation,
\begin{equation}
    \label{2.21}
    \begin{split}
        \{\hat{b}_{h}(\vec{k}),\hat{b}^{\dagger}_{h'}(\vec{k}')\} &= (2\pi)^{D-1}\delta^{D-1}(\vec{k} - \vec{k}')\delta_{hh'}\\
        \{\hat{b}_{h}(\vec{k}),\hat{b}_{h'}(\vec{k}')\} &=0\,,
        \hspace{1.8cm}
        \{\hat{b}^{\dagger}_{h}(\vec{k}),\hat{b}^{\dagger}_{h'}(\vec{k}')\} = 0.\hspace{3.8cm}
    \end{split}
\end{equation}
$A_{h,\alpha}(\eta, \vec{k})$ and $\bar A_{h,\alpha}(\eta, \vec{k})$ in
Eqs.~\eqref{2.18} and~(\ref{2.18b}) are defined as,
\begin{equation}
\begin{split}
 \label{2.22}
		   A_{h,\alpha}(\eta, \vec{k}) &= \begin{pmatrix}
		    \rho_{h}(\eta, \vec{k})\\
		    \epsilon\rho^{*}(\eta, -\vec{k})
		    \end{pmatrix} = \begin{pmatrix}
		    L_{h}(\eta, \vec{k})\xi_{h}(\Vec{k})\\
		    L^{*}_{h}(\eta, -\vec{k})\epsilon\xi^{*}_{h}(-\Vec{k})
		    \end{pmatrix},
		     \\
 \bar A_{h,\alpha}(\eta, \vec{k}) &= \begin{pmatrix}-\rho^{T}_{h}(\eta,-\vec{k})\epsilon & \rho^{\dagger}_{h}(\eta,\vec{k})\end{pmatrix}  =
		 \begin{pmatrix}- L_{h}(\eta,-\vec{k})\xi_{h}^{T}(-\vec{k}\,)\epsilon \;&\;  L_{h}(\eta,-\vec{k})\xi_{h}^{\dagger}(\vec{k}\,)\end{pmatrix}
\,,
\end{split}	
\end{equation}
 where $\rho_{h}(\eta, \vec{k})$ is the rescaled helicity eigenspinor in momentum space,  
 $L_{h}(\eta, \vec{k})$ is the Majorana particle mode function, and $\xi_{h}(\Vec{k})$ 
 is the helicity eigenspinor defined by,
 \begin{equation}
     \label{2.23}
     \hat{h}\xi_{h}(\vec{k}) = h\xi_{h}(\vec{k})\hspace{1cm};\hspace{2cm}h=\pm1,
 \end{equation}
 where $\hat{h}$ is the helicity operator, which in $D=4$ has the simple form, 
 $\hat{h}=(\vec{k}/\|\vec{k}\|)\cdot\vec{\sigma}$
%and this relation holds in $D$ dimensions as well
\footnote{We have defined the  Majorana fields in D dimensions and the chirality operators can be readily generalized to higher dimensions as was carefully done in Ref.~\cite{Koksma:2009tc}. 
The properties listed for the helicity $2$-spinor in Appendix~\eqref{Appendix-A} 
Eqs.~(\ref{A.5}--\ref{A.7}) and Eq.~\eqref{A.10} stay the same in general $D$ dimensions. Thus, the definitions in $D=4$ that we use in this section for the construction of the propagator remain valid in $D$ dimensions.} and using Eq.~\eqref{A.5} we can write $A_{h}(\eta, \vec{k})$ and $\bar{A}_{h}(\eta, \vec{k})$, 
\begin{equation}
    \label{2.31}
    \begin{split}
  A_{h}(\eta, \vec{k}) &= 
     \begin{pmatrix}
    L_{h}(\eta, \vec{k}\,)\\
   h e^{i\theta(\hat{k}_{x},\hat{k}_{y})}L^{*}_{h}(\eta, -\vec{k}\,)
    \end{pmatrix}\otimes\xi_{h}(\vec{k}\,)
 \\
  \bar{A}_{h}(\eta, \vec{k}) &= \begin{pmatrix}
  h  e^{-i\theta(\hat{k}_{x},\hat{k}_{y})}L_{h}(\eta, -\vec{k}\,) \;&\; L^{*}_{h}(\eta, \vec{k}\,)
    \end{pmatrix}\otimes\xi^{\dagger}_{h}(\vec{k}\,)
\,,
    \end{split}
\end{equation}
where  $\theta(\hat{k}_{x},\hat{k}_{y})= \tan^{-1}\big(\hat k_y/\hat k_x\big)$
(see Eq.~(\ref{A.4})). 

In order to progress towards normalization of the Majorana mode functions, 
note that the following identity for the chiral eigenvectors holds, 
\begin{equation}
    \label{2.25a}
    \sum_{h=\pm}\left(\xi^{\dagger}_{h,a}(\vec{k})\otimes\xi_{h,b}(\vec{k})\right) = \delta_{ab}\hspace{0.4cm},\hspace{1cm} 
    \big(-\infty<\ k_{i} <+\infty\hspace{0.2cm};\hspace{0.2cm} i\in\{x,y,z\}
    \big)\,,
\end{equation}
which is proved  in Appendix~C.
Next, since the helicity eigenstates $ L_{\pm}(\eta, \vec{k})$ do not couple to the vacuum, 
their normalization should be independent on helicity, 
\begin{equation}
    \label{2.26}
    | L_{+}(\eta, \vec{k}) |^{2} = | L_{-}(\eta, \vec{k}) |^{2}.
\end{equation}
Making use of Eqs.~(\ref{2.20}--\ref{2.21}) 
along with the definitions in Eq.~\eqref{2.18} and~\eqref{2.19}, one can see that the normalisation condition is given by,
\begin{equation}
    \label{2.25}
    \sum_{h=\pm}\left[| L_{h}(\eta, \vec{k})|^{2}\left(\xi^{\dagger}_{h,a}(\vec{k})\otimes\xi_{h,b}(\vec{k})\right)\right] = \delta_{ab}\hspace{0.4cm},\hspace{1cm} 
    \big(-\infty<\ k_{i} <+\infty\hspace{0.2cm};\hspace{0.2cm} i\in\{x,y,z\}
    \big)\,.
\end{equation}
Finally, from Eqs.~\eqref{2.25a} and~\eqref{2.26} from the above equation it follows that,
\begin{equation}
    \label{2.27}
 %   \begin{split}
        | L_{\pm}(\eta, \vec{k}) |^{2} = 1%\frac{1}{2}
        \hspace{1cm} 
        \left(-\infty<k_{i}<\infty\hspace{0.2cm}; \hspace{0.2cm} i\in\{x,y,z\} \right)
\,.
%\\
%        | L_{-}(\eta, \vec{k}) |^{2} = \frac{1}{2}\hspace{1cm} 
 %       \left(-\infty<k_{i}<\infty\hspace{0.2cm}; \hspace{0.2cm} i\in\{x,y,z\}  \right).
%    \end{split}
\end{equation}
This normalization differs from the usual one for the Dirac fermions, in which the mode functions squared
are normalized to $1/2$, see {\it e.g.} Ref.~\cite{Koksma:2009tc},\cite{Garbrecht:2006jm}. This difference can be traced back 
to the factor $2$ in the anticommutation relation~(\ref{2.20}).
Eq.~\eqref{2.26} can then be written as follows,
\begin{equation}
    \label{2.28}
     \sum_{h=\pm}\left[| L_{h}(\eta, \vec{k})|^{2} 
     + | L_{h}(\eta, -\vec{k})|^{2}\right] = 2 %1 
     \hspace{0.8cm} \hspace{1cm} \left( 0<k_{i}<\infty\hspace{0.2cm}; \hspace{0.2cm} i\in\{x,y,z\}  \right)
     \,.
\end{equation}
Here it is important to notice that, due to the hemispherical construction of helicity eigenmodes (discussed in detail in section~\ref{Mode function solutions} below),
for a given helicity and momentum $(h=\pm,\vec k)$, 
only one of the mode function densities 
contributes (the other is zero). For example, for $h=+1$, $| L_{+}(\eta, \vec{k})|^{2}$ contributes and $| L_{-}(\eta, \vec{k})|^{2}$ is zero. This observation is of 
an essential importance for the correct normalization of the Majorana 
propagator, which is one property that can be used to distinguish between the Dirac and Majorana particles.

\subsubsection{Mode function solutions}
\label{Mode function solutions}

We can now describe the equations of motion for Majorana particles. For this we use the Dirac equation given by 
Eq.~\eqref{2.10} which gives the following equations of motion for the mode functions for $k_{z}\geq0$:
\begin{equation}
    \label{2.29}
    \begin{split}
        i\partial_{\eta}L^{*}_{h}(\eta, -\vec{k}) - h\|\vec{k}\| L^{*}_{h}(\eta, -\vec{k}) 
           - ahm^{*}L_{h}(\eta, \vec{k})e^{-i\theta} &= 0\\
        i\partial_{\eta}L_{h}(\eta, \vec{k}) + h\| \vec{k} \| L_{h}(\eta, \vec{k})
         - ahmL^{*}_{h}(\eta, -\vec{k})e^{i\theta} &= 0
\,.
    \end{split}
\end{equation}
Upon writing $h=e^{i\frac{\pi}{2}\left(1-h\right)} = e^{-i\frac{\pi}{2}\left(1-h\right)}$ and $m = |m| e^{i\phi}$, 
Eq.~\eqref{2.29} becomes,
\begin{equation}
    \label{2.30}
      \begin{split}
      ie^{\frac{i\bar{\phi}}{2}}\partial_{\eta}L^{*}_{h}(\eta, -\vec{k}) - e^{\frac{i\bar{\phi}}{2}}h\| \vec{k}\| L^{*}_{h}(\eta, -\vec{k}) -e^{-\frac{i\bar{\phi}}{2}}a| m| L_{h}(\eta, \vec{k}) &= 0\\
      ie^{-\frac{i\bar{\phi}}{2}}\partial_{\eta}L_{h}(\eta, \vec{k}) + e^{-\frac{i\bar{\phi}}{2}}h\| \vec{k}\| L_{h}(\eta, \vec{k}) -e^{\frac{i\bar{\phi}}{2}}a| m | L^{*}_{h}(\eta, -\vec{k}) &= 0 ,
  \end{split} 
\end{equation}
where we have defined $\bar{\phi} = \theta(\hat{k}_{x},\hat{k}_{y}) + \frac{\pi}{2}\left(h - 1\right) + \phi$.
%and redefined the mass matrix as $\Tilde{M} = a| m | \mathbb{1}_{2\times 2}$.
 It is convenient to transform to the positive/negative frequency basis,
\begin{equation}
    \label{2.32}
   u_{h\pm}(\eta, \vec{k}\,) = \frac{1}{\sqrt{2}}\left(e^{-\frac{i\bar{\phi}}{2}}L_{h}(\eta, \vec{k}\,) \pm e^{\frac{i\bar{\phi}}{2}}L^{*}_{h}(\eta, -\vec{k}\,)\right)
\,,
\end{equation}
in which the mode functions $u_{h\pm}$ obey a Bessel's differential equation, 
\begin{equation}
    \label{2.33}
    \partial^{2}_{\eta}u_{h\pm}(\eta, \vec{k}\,) + \left(\| \vec{k} \|^{2} + \frac{\frac{1}{4} - \left(\frac{1}{2}\mp\frac{i| m |}{H}\right)^{2}}{\eta^{2}}\right)u_{h\pm}(\eta, \vec{k}\,) = 0.
\end{equation}
The Bunch-Davies vacuum solutions~\footnote{By judiciously introducing Bogolyubov coefficients, 
one can quite straightforwardly
construct positive and negative frequency mode functions for more general (pure) Gaussian states. 
Details of such a construction for Dirac fermions
can be found in Ref.~\cite{Koksma:2009tc}. Here we are primarily interested in the vacuum propagator 
in de Sitter, for which the Bunch-Davies choice~\eqref{2.34} sufficies.}  can be written in terms of Hankel functions 
as follows ({\it cf.} Ref.~\cite{Garbrecht:2006jm}),
\begin{equation}
    \label{2.34}
    \begin{split}
        u_{h+}(\eta, \vec{k}) &= %\frac{1}{\sqrt{2}}
         e^{\frac{i\pi\nu_{+}}{2}}\sqrt{-\frac{\pi\| \vec{k} \|\eta}{4}}H^{(1)}_{\nu_{+}}(-\| \vec{k}\|\eta)\\
        u_{h-}(\eta, \vec{k}) &= %\frac{h}{\sqrt{2}} 
        he^{\frac{i\pi\nu_{-}}{2}}\sqrt{-\frac{\pi\| \vec{k} \|\eta}{4}}H^{(1)}_{\nu_{-}}(-\| \vec{k}\|\eta)
\,.
    \end{split}
\end{equation}
Here $H^{(1)}_{\nu}(-\|\vec{k}\|\eta)$ is Hankel function of the first kind, which in the ultraviolet (equivalently in a distant past, $k\eta\rightarrow -\infty$), reduces to the positive frequency vacuum solution. The indices $\nu_{\pm}$ are given by,
\begin{equation}
    \label{2.35}
   % \begin{split}
        \nu_{\pm} = \frac{1}{2} \mp i\zeta
\,,\qquad
        \zeta = \frac{| m |}{H}
\,.\qquad
    %\end{split}
\end{equation}
We can then write the set of equations in Eq.~\eqref{2.30} for $k_{z}\leq0$ as follows,
\begin{equation}
    \label{2.36}
    \begin{split}
      ie^{\frac{i\bar{\phi}}{2}}\partial_{\eta}L^{*}_{h}(\eta, \vec{k}\,) 
      - e^{\frac{i\bar{\phi}}{2}}h\| \vec{k}\| L^{*}_{h}(\eta, \vec{k}\,) 
      +  e^{-\frac{i\bar{\phi}}{2}}a| m | L_{h}(\eta, -\vec{k}\,) &= 0\\
      ie^{-\frac{i\bar{\phi}}{2}}\partial_{\eta}L_{h}(\eta, -\vec{k}\,) 
      + e^{-\frac{i\bar{\phi}}{2}}h\| \vec{k}\| L_{h}(\eta, -\vec{k}\,) 
      +  e^{\frac{i\bar{\phi}}{2}}a| m | L^{*}_{h}(\eta, \vec{k}\,) &= 0
\,,
  \end{split}
\end{equation}
for which we again transform into the positive/negative frequency basis,
\begin{equation}
\label{2.37}
    u_{h\pm}(\eta, -\vec{k}) = \frac{1}{\sqrt{2}}\left(L_{h}(\eta, -\vec{k})e^{-\frac{i\bar{\phi}}{2}}\pm L^{*}_{h}(\eta, \vec{k})e^{\frac{i\bar{\phi}}{2}}\right),
\end{equation}
where 
\vskip -0.5cm
\begin{equation}
    \label{2.38}
    \begin{split}
    u_{h+}(\eta, -\vec{k}\,) 
    &= %\frac{1}{\sqrt{2}}
    e^{-\frac{i\pi\nu_{-}}{2}}\sqrt{-\frac{\pi\| \vec{k} \|\eta}{4}}
       H^{(2)}_{\nu_{-}}(-\| \vec{k}\|\eta)
\\
    u_{h-}(\eta, -\vec{k}\,) &= %-\frac{h}{\sqrt{2}} 
    -he^{\frac{-i\pi\nu_{+}}{2}}\sqrt{-\frac{\pi\| \vec{k} \|\eta}{4}}
    H^{(2)}_{\nu_{+}}(-\| \vec{k}\|\eta)
\,.
    \end{split}
\end{equation}
 Here $H^{(2)}_{\nu}(-\|\vec{k}\eta\|)$ is Hankel function of the second kind, which in the ultraviolet regime reduces to the negative frequency vacuum solution. 
 From Eq.~(\ref{2.28}) one can infer that this set of solutions satisfies the following normalization condition,
\begin{equation}
    \label{2.39}
    \begin{split}
        |u_{h+}(\eta, \vec{k})|^{2} + |u_{h-}(\eta, \vec{k})|^{2} 
        + |u_{h+}(\eta, -\vec{k})|^{2} + |u_{h-}(\eta, -\vec{k})|^{2} = 2
\,.
    \end{split}
\end{equation}
From Eq.~\eqref{2.34}) and Eq.~\eqref{2.38} one finds a complete set of solutions for the mode functions,
\begin{eqnarray}
    \label{2.40}
    e^{-\frac{i\bar{\phi}}{2}}L_{h}(\eta, \vec{k}) 
    \!\!&=&\!\! %\frac{1}{2}
    \sqrt{-\frac{\pi\| \textbf{k}\|\eta}{8}}\left[e^{\frac{i\pi\nu_{+}}{2}}H^{(1)}_{\nu_{+}}(-\|\vec{k}\|\eta) + he^{\frac{i\pi\nu_{-}}{2}}H^{(1)}_{\nu_{-}}(-\|\vec{k}\|\eta)\right],
\\
    \label{2.41}
    e^{\frac{i\bar{\phi}}{2}}L^{*}_{h}(\eta, -\vec{k})
    \!\!&=&\!\! %\frac{1}{2}
    \sqrt{-\frac{\pi\| \vec{k}\|\eta}{8}}\left[e^{\frac{i\pi\nu_{+}}{2}}H^{(1)}_{\nu_{+}}(-\|\textbf{k}\|\eta) - he^{\frac{i\pi\nu_{-}}{2}}H^{(1)}_{\nu_{-}}(-\|\vec{k}\|\eta)\right],
\\
    \label{2.42}
    e^{-\frac{i\bar{\phi}}{2}}L_{h}(\eta, -\vec{k})
    \!\!&=&\!\!  %\frac{1}{2}
    \sqrt{-\frac{\pi\| \vec{k}\|\eta}{8}}\left[e^{\frac{-i\pi\nu_{-}}{2}}H^{(2)}_{\nu_{-}}(-\|\vec{k}\|\eta) - he^{\frac{-i\pi\nu_{+}}{2}}H^{(2)}_{\nu_{+}}(-\|\vec{k}\|\eta)\right],
\\
    \label{2.43}
    e^{\frac{i\bar{\phi}}{2}} L^{*}_{h}(\eta, \vec{k}) 
    \!\!&=&\!\! 
    \sqrt{-\frac{\pi\| \vec{k}\|\eta}{8}}
    \left[e^{\frac{-i\pi\nu_{-}}{2}}H^{(2)}_{\nu_{-}}(-\|\vec{k}\|\eta) 
    + h{\rm e}^{\frac{-i\pi\nu_{+}}{2}}H^{(2)}_{\nu_{+}}(-\|\vec{k}\|\eta)\right].
\end{eqnarray}

\begin{figure}[h]
 \vskip -1cm
    %\centering
   {{\includegraphics[width=0.4\textwidth]{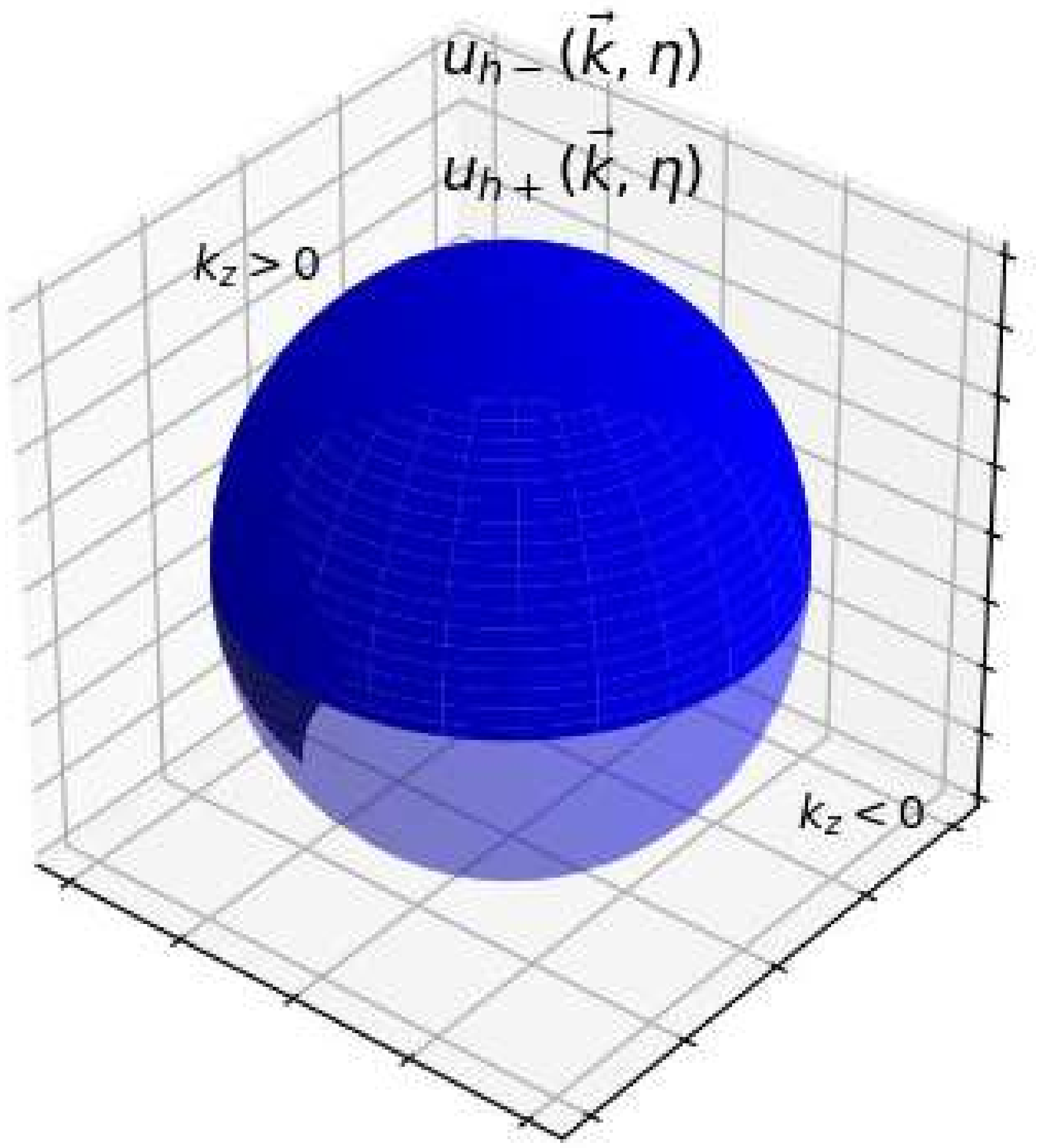} }}%
    \qquad
 {{\includegraphics[width=0.4\textwidth]{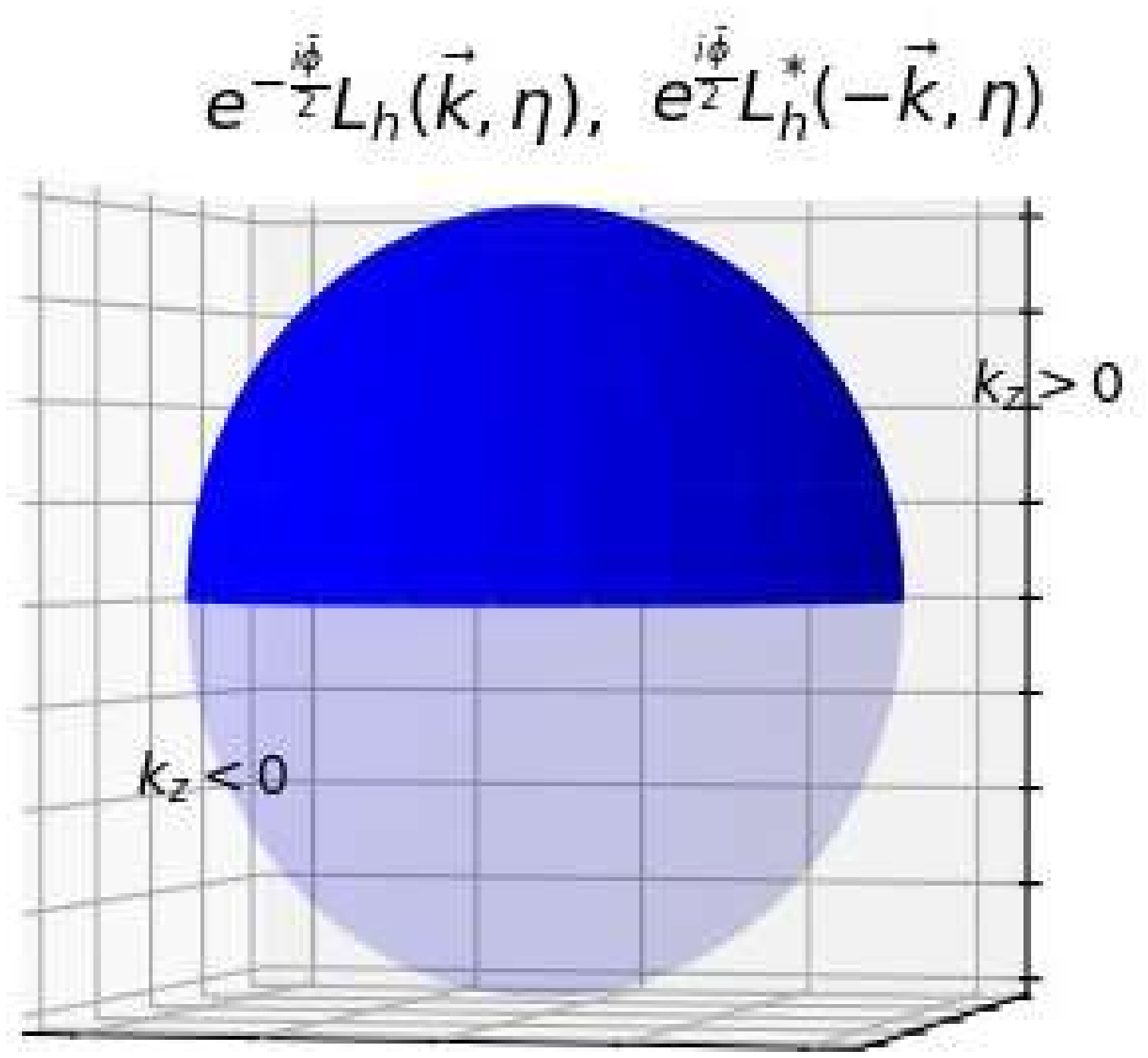} }}
 %{{\includegraphics[width=0.4\textwidth]{Fig3_3d.jpg} }}
 \vskip -1.7cm
   \caption{
   {\it Left panel} shows the positive and negative frequency solutions $u_{h\pm}(\eta, \vec{k})$,
   defined on the upper hemisphere 
 ($k_{z}>0$) and given by 
$u_{h+}(\eta, \vec{k}) = e^{\frac{i\pi\nu_{+}}{2}}
 \sqrt{-\frac{\pi\|\vec{k}\|\eta}{4}}H^{(1)}_{\nu_{+}}(-\|\vec{k}\|\eta)$,
    $u_{h-}(\eta, \vec{k}) = h
    e^{\frac{i\pi\nu_{-}}{2}}
    \sqrt{-\frac{\pi\|\vec{k}\|\eta}{4}}H^{(1)}_{\nu_{-}}(-\|\vec{k}\|\eta)$. 
    Hence the solutions to the corresponding mode functions {(\it right panel)} are also defined on the 
   upper hemisphere.
    }
   \label{fig:Fig1} 
\end{figure}
\begin{figure}
\vskip -1.5cm
    \centering
   {{\includegraphics[width=0.4\textwidth]{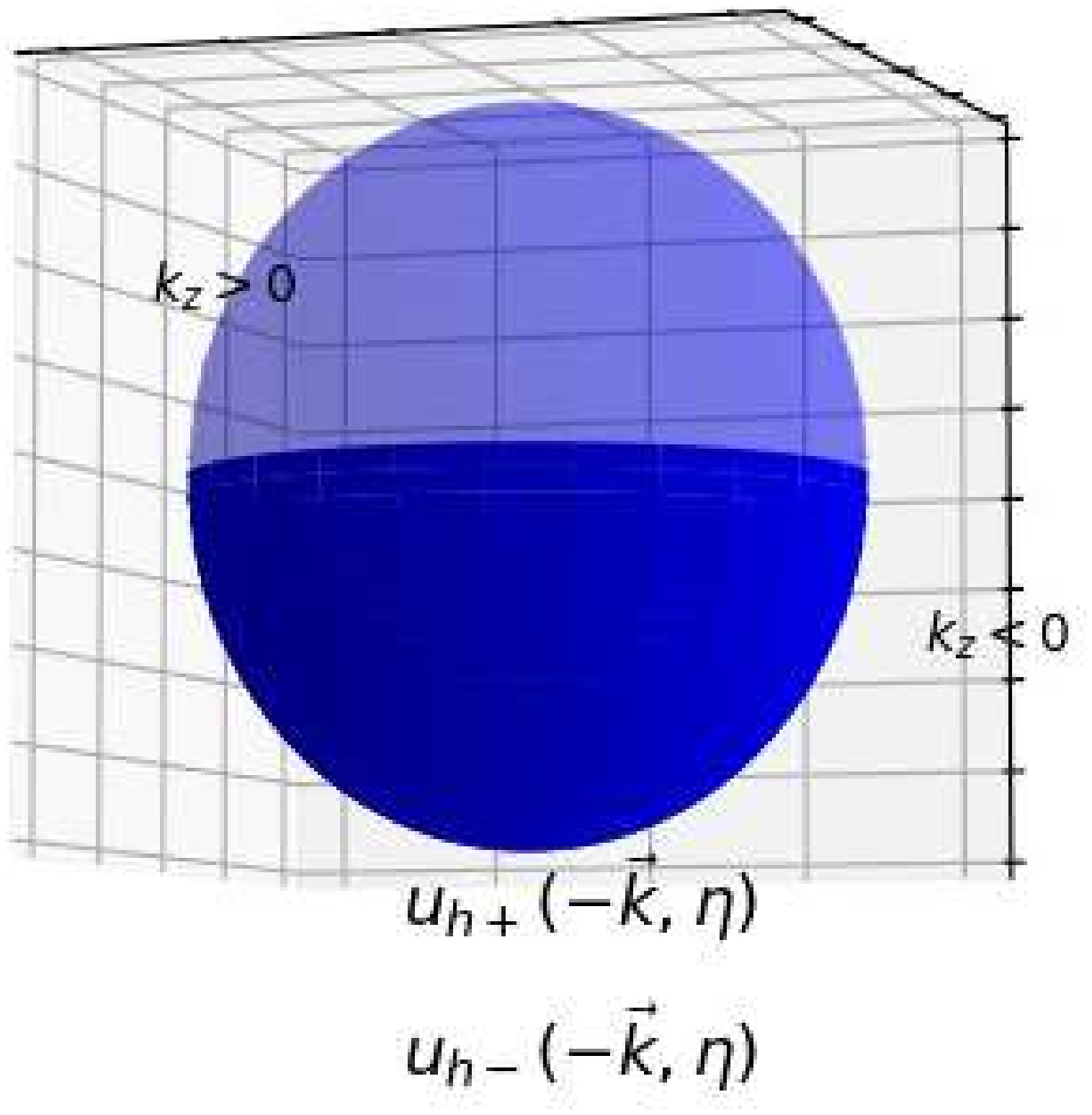} }}%
    \qquad
  {{\includegraphics[width=0.4\textwidth]{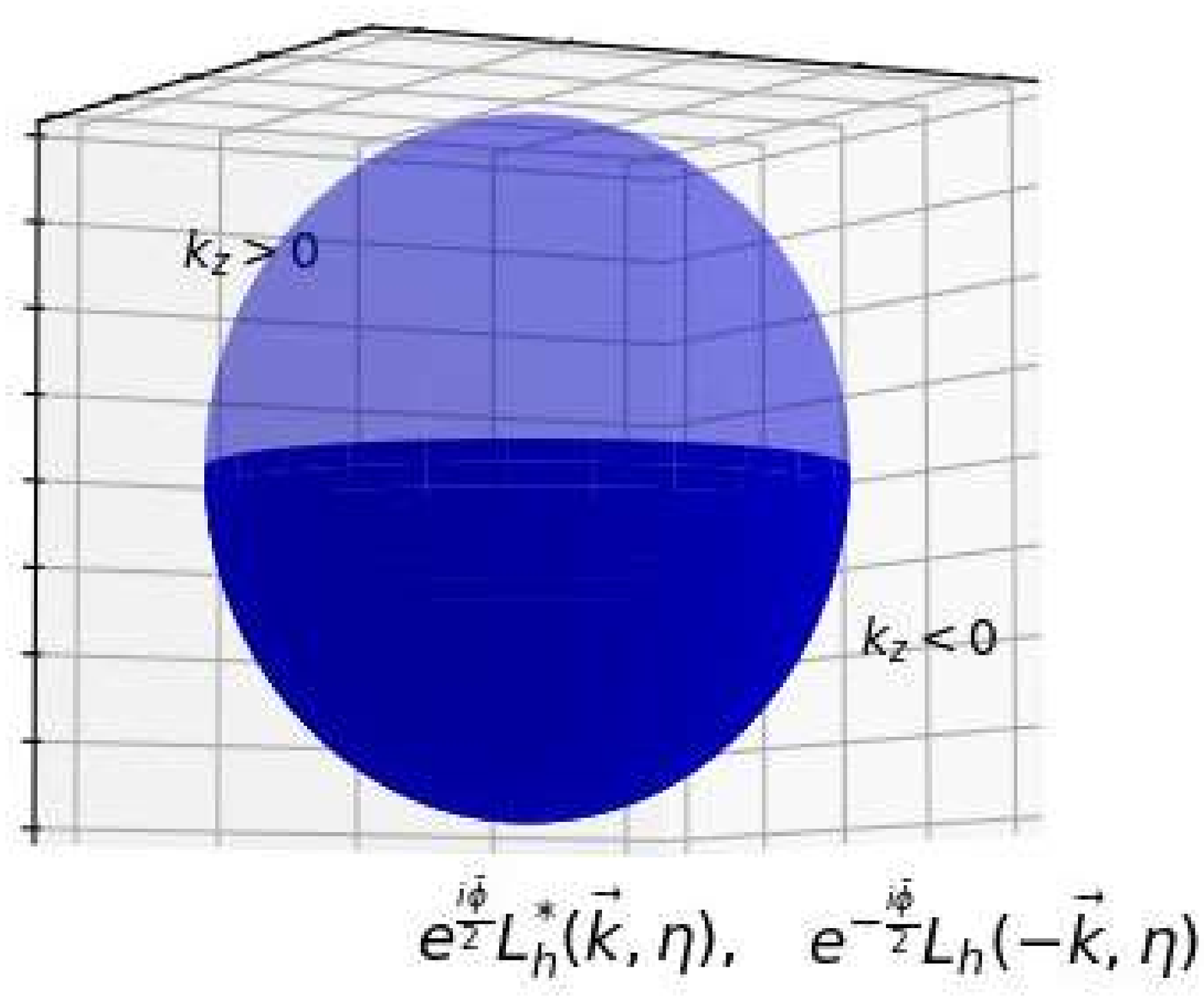} }}%
\vskip -1.8cm
    \caption{
    {\it Left panel} shows the positive and negative frequency solutions 
    $u_{h\pm}(\eta, -\vec{k})$ defined on the lower hemisphere ($k_{z}<0$) and
    given by $u_{h+}(\eta, -\vec{k})
     = e^{\frac{i\pi\nu_{-}}{2}}
    \sqrt{-\frac{\pi\|\vec{k}\|\eta}{4}}H^{(2)}_{\nu_{-}}(-\|\vec{k}\|\eta)$, 
    $u_{h-}(\eta, -\vec{k}) = -he^{\frac{i\pi\nu_{+}}{2}}
    \sqrt{-\frac{\pi\|\vec{k}\|\eta}{4}}H^{(2)}_{\nu_{+}}(-\|\vec{k}\|\eta)$.
    {\it Right panel} depicts the solutions to the mode functions, which are also defined 
    on the lower hemisphere.
    } 
   \label{fig:Fig2}
\end{figure}

\noindent
Solutions in Eqs.~\eqref{2.40}--\eqref{2.41} are valid for $k_{z}\geq0$ 
and in Eqs.~\eqref{2.42}--\eqref{2.43} are valid for $k_{z}\leq0$. To see this, we can look at 
Fig.~\ref{fig:Fig1} (left panel) where we see that on the upper hemisphere, 
the solutions are described by $u_{h\pm}(\eta, \vec{k})$, which are given in terms of Hankel functions of the first kind $H^{(1)}_{\nu}(-\|\vec{k}\|\eta)$ and thus, the solutions for $e^{-\frac{i\bar{\phi}}{2}}L_{h}(\eta, \vec{k})$ and $e^{\frac{i\bar{\phi}}{2}}L^{*}_{h}(\eta, -\vec{k})$ are also defined on this hemisphere, 
as can be seen in Fig.~\ref{fig:Fig1} (right panel). 
However, when we consider $\vec{k}\rightarrow-\vec{k}$, then the solutions are defined on the lower hemisphere, 
shown in Fig.~\ref{fig:Fig2} (left panel), and described by $u_{h\pm}(\eta, -\vec{k})$, which 
in turn are given 
in terms of Hankel functions of the second kind $H^{(2)}_{\nu}(-\|\textbf{k}\|\eta)$,
 and therefore the solutions for $e^{-\frac{i\bar{\phi}}{2}}L_{h}(\eta, -\vec{k})$ and $e^{\frac{i\bar{\phi}}{2}}L^{*}_{h}(\eta, \vec{k})$ are also defined on the lower hemisphere, which is illustrated in Figure~\ref{fig:Fig2} (right panel).  

%%%%%%%%%%%%%%%%%%%%%%%%%%
%%%     M A J O R A N A    P R O P A G A T O R   %%%
%%%%%%%%%%%%%%%%%%%%%%%%%%

\section{Majorana Propagator}
\label{Majorana Propagator}
\label{Section-3}

Before we begin with construction of the Majorana propagator, let us define some important quantities 
for de Sitter space. Firstly, the scale factor on the expanding (Poincar\'e) patch of de Sitter space reads,
\begin{equation}
    \label{3.1}
    a(\eta) = - \frac{1}{H\eta}\hspace{1cm}
    \big(\eta<0,\ H = \text{const.}\big)
\,.
\end{equation}
Next, the following invariant distance (biscalar) functions are useful,
\begin{eqnarray}
    \label{3.2}
    y_{++}(x;x') \!\!& =&\!\! \frac{\Delta x^{2}_{++}(x;x')}{\eta\eta'}
        = \frac{1}{\eta\eta'}\left(- \left(\lvert \eta - \eta'\rvert - i\varepsilon\right)^{2} + \|\Delta\vec{x} \|^{2} \right)
\\
    \label{3.3}
    y_{+-}(x;x') \!\!& =&\!\! \frac{1}{\eta\eta'}\left(- \left(\eta - \eta' + i\varepsilon\right)^{2} + \| \Delta\vec{x} \|^{2}\right)
\\
    \label{3.4}
    y_{-+}(x;x') \!\!& =&\!\! \frac{1}{\eta\eta'}\left(- \left(\eta - \eta' - i\varepsilon\right)^{2} + \| \Delta\vec{x} \|^{2}\right)
\\
    \label{3.5}
    y_{--}(x;x')\!\!& =&\!\! \frac{1}{\eta\eta'}\left(- \left(\lvert \eta - \eta'\rvert + i\varepsilon\right)^{2} +  \|\Delta\vec{x} \|^{2}\right),
\end{eqnarray}
where $i\varepsilon$ denotes an imaginary infinitesimal time shift ($\epsilon\ll 1$),
which are useful for definition of  various 
two-point functions on de Sitter. Eqs.~(\ref{3.1}--\ref{3.4}) imply the following identities,
\begin{equation}
    \label{3.6}
    \begin{split}
        y_{++}(x;x') &= \Theta(\eta - \eta')y_{-+}(x;x') + \Theta(\eta' - \eta)y_{+-}(x;x')\\
        y_{--}(x;x') &= \Theta(\eta - \eta')y_{+-}(x;x') + \Theta(\eta' - \eta)y_{-+}(x;x')
        \,.
    \end{split}
\end{equation}
These distance functions are related to the geodesic distance on de Sitter space $\ell(x;x')$ as follows, 
$4\sin^2[H\ell(x;x')/2]= y_{ab}(x;x')|_{\epsilon\rightarrow 0}\; (a,b=\pm)$.

\subsection{Construction of the propagator}
\label{Construction of the propagator}

The Majorana propagator obeys the equation, 
\begin{equation}
  \sqrt{-g}\left[i\gamma^{\mu}\nabla_{\mu}-{\cal M}\right]iS_F(x;x') = i \delta^D(x-x')
\,,
    \label{3.7a}
\end{equation}
and it ought to be symmetric under the exchange of $x$ and $x'$.
To solve for the propagator on de Sitter space, we shall construct the propagator by inserting 
the mode function decomposition~(\ref{2.18}) into the definition of the propagator,
\begin{eqnarray}
    \label{3.7}
        iS_{\alpha\beta}(x;x') \!\!&=&\!\! a^{-\frac{D-1}{2}}(\eta)a^{-\frac{D-1}{2}}(\eta')i\tilde S_{\alpha\beta}(x;x')
\,,\quad
\\
    \label{3.7b}
           i\tilde S_{\alpha\beta}(x;x') \!\!&=&\!\! \Big\langle \Omega \Big\lvert T\Big[\hat{\Tilde{\Psi}}_{M,\alpha}(x)
       \hat{\Tilde{\bar{\Psi}}}_{M,\beta}(x')\Big] \Big\rvert \Omega \Big\rangle
       \hspace{7.5cm}\\
\nonumber
         \!\!&=&\!\! \Theta(\eta \!-\! \eta')\Big\langle \Omega\Big \lvert\hat{\Tilde{\Psi}}_{M,\alpha}(x)\hat{\bar{\Tilde{\Psi}}}_{M,\beta}(x')\Big\rvert \Omega \Big\rangle 
         - \Theta(\eta' \!-\! \eta) \Big\langle \Omega \Big\lvert 
        \hat{\bar{\Tilde{\Psi}}}_{M,\beta}(x')\hat{\Tilde{\Psi}}_{M,\alpha}(x)
         \Big\rvert \Omega \Big\rangle
.\quad
\end{eqnarray}
This results in,
%
%\begin{equation}
  %  \label{3.8}
   % \begin{split}
   %     i\tilde S_{\alpha\beta}(x,x') = \Theta(\eta \!-\! \eta')\sum_{h,h' = \pm}\int \frac{d^{D-1}k}{(2\pi)^{D-1}}\frac{d^{D-1}k'}{(2\pi)^{D-1}}e^{i\vec{k}\cdot\vec{x} 
 %       - i\vec{k}'\cdot\vec{x}'}\langle \hat{b}_{h}(\vec{k})\hat{b}^{\dagger}_{h'}(-\vec{k}')\rangle A_{h,\alpha}(\eta, \vec{k})\bar{A}_{h',\beta}(\vec{k}',\eta') 
  %      \\- \Theta(\eta' \!-\! \eta)\sum_{h,h' = \pm}\int \frac{d^{D-1}k}{(2\pi)^{D-1}}\frac{d^{D-1}k'}{(2\pi)^{D-1}}e^{-(i\vec{k}\cdot\vec{x} - i\vec{k}'\cdot\vec{x}')}\langle \hat{b}_{h}(\vec{k})\hat{b}^{\dagger}_{h'}(-\vec{k}')\rangle A_{h,\alpha}(\eta, -\vec{k})\bar{A}_{h',\beta}(-\vec{k}',\eta').
 %   \end{split}
%\end{equation}
%
 %we can then write this as follows
\begin{equation}
    \label{3.9}
    \begin{split}
        i\tilde S_{\alpha\beta}(x;x') 
        = \Theta(\eta \!-\! \eta')\sum_{h=\pm}\int \frac{d^{D-1}k}{(2\pi)^{D-1}}e^{i\vec{k}\cdot(\vec{x} - \vec{x}')}
         A_{h,\alpha}(\eta, \vec{k})\bar{A}_{h,\beta}(\eta', \vec{k}) \\
        - \Theta(\eta' \!-\!  \eta)\sum_{h=\pm}\int\frac{d^{D-1}k}{(2\pi)^{D-1}}e^{-i\vec{k}\cdot(\vec{x} - \vec{x}')}
         A_{h,\alpha}(\eta, -\vec{k})\bar{A}_{h,\beta}(\eta', -\vec{k}),
    \end{split}
\end{equation}
where $A_{h,\alpha}(\eta, \vec{k})$ and $\bar A_{h,\beta}(\eta', \vec{k})$ are defined in
Eq.~\eqref{2.31}
and we have used the anti-commutation relation in Eq.~\eqref{2.21} 
(with the help of which we have performed 
one of the momentum integrals).

Further steps in the construction of the propagator are analogous to those in Ref.~\cite{Koksma:2009tc}. 
The propagator is computed, component-by-component, using the solutions for the mode functions 
from Eqs.~(\ref{2.40}--\ref{2.43}). The propagator components are defined in terms of
the projection operators using the following identity,
\begin{equation}
    \label{3.10}
    \left(i\gamma^{b}\partial_{b} + a|m|\right) \frac{1\pm\gamma^{0}}{2} = \frac{1}{2}\begin{pmatrix}
    \pm i\partial_{\eta} \pm i\sigma^{i}\partial_{i} + a| m |  &  i\partial_{\eta} + i\sigma^{i}\partial_{i} \pm | m |\\
    i\partial_{\eta} - i\sigma^{i}\partial_{i} \pm a | m | & \pm i\partial_{\eta} \mp i\sigma^{i}\partial_{i} + a| m |
    \end{pmatrix}.
\end{equation}
After several steps,~\footnote{For details we refer to Ref.~\cite{Koksma:2009tc}.} 
and in particular upon making use of Eq.~\eqref{2.7}, one arrives 
at the Majorana propagator in the form, 
\begin{equation}
    \label{3.11A}
\left[i S_F(x;x')\right]^{\tt rot}
  =  a\left(i\gamma^{\!\mu}\nabla_{\mu} \!+\!|m|\right)\frac{H^{D-2}}{\sqrt{aa'}}\left[\frac{1 +\gamma^{0}}{2}iS^{+}(x;x') \!+\! \frac{1-\gamma^{0}}{2}i\Tilde{S}^{+}(x;x')\right],
\end{equation}
where $H^{D-2} = \left(aa'\eta\eta'\right)^{\frac{2-D}{2}}$ and $H$ is the Hubble rate,
which is constant on de Sitter. The functions 
$iS^{+}(x;x')$ and $i\Tilde{S}^{+}(x;x')$ are biscalars defined as follows,
\begin{equation}
    \label{3.12}
    \begin{split}
        iS^{+}(x;x') &= \frac{\Gamma\left(\frac{D}{2} \!+\! i\zeta\right)\Gamma\left(\frac{D-2}{2} \!-\!  i\zeta\right)}{(4\pi)^{D/2}\Gamma(\frac{D}{2})}
        \times{}_{2}F_{1}\left(\frac{D}{2}\!+\!i\zeta,\frac{D-2}{2}\!-\! i\zeta;\frac{D}{2};1-\frac{y_{++}(x;x')}{4}\right)
\\
        i\Tilde{S}^{+}(x;x') &= \frac{\Gamma\left(\frac{D}{2}\!-\! i\zeta\right)\Gamma\left(\frac{D-2}{2} \!+\! i\zeta\right)}{(4\pi)^{D/2}\Gamma(\frac{D}{2})}
        \times{}_{2}F_{1}\left(\frac{D}{2}\!-\! i\zeta,\frac{D-2}{2}\!+\!i\zeta;\frac{D}{2};1\!-\! \frac{y_{++}(x;x')}{4}\right)
\,,
    \end{split}
\end{equation}
where ${}_2F_1$ denote Gauss' hypergeometric functions and we have used the definitions listed in Eq.~\eqref{3.6}. To calculate the integral we have also made use of the properties of Hankel functions in Appendix~D (the details of the calculation can be found in an Appendix B of~\cite{Koksma:2009tc}, Eq.~(B.1) and~(B.4--B.6)) and the relevant integrals needed for integration of products of Bessel functions can be found in identity (6.578.10) of Ref.~\cite{Gradshteyn:2014}.

The propagator~(\ref{3.11A}) is still in the basis in which the phase 
$\phi = {\rm Arctan}[m_2/m_1]\;$ ($m=m_1+im_2$) of the mass term is removed,
and it is the propagator for Majorana fermions on de Sitter whose mass is real.
When the mass is complex however,  one ought to bring it back to the standard form 
by performing an additional chiral rotation,
\begin{equation}
    iS_F(x;x') = %{\rm e}^{-\frac{i\phi}{2}}
               {\rm e}^{-\frac{i\phi}{2}\gamma^{5}} 
            \left[i S_F(x;x')\right]^{\tt rot} {\rm e}^{-\frac{i\phi}{2}\gamma^{5}}       
               %    {\rm e}^{\frac{i\phi}{2}}
\,.
\label{chiral rotation}
\end{equation}
%
%where $\phi = {\rm Arctan}[m_2/m_1]$ is the phase of the complex mass $m$.
When commuted through the kinetic operator in~(\ref{3.11A}), 
this then gives,  
\begin{equation}
    \label{3.11}
    iS_{F}(x;x') = a\left(i\gamma^{\!\mu}\nabla_{\mu} +{\cal M}^\dagger\right)\frac{H^{D-2}}{\sqrt{aa'}}\left[\frac{1 \!+\! {\rm e}^{i\phi\gamma^5}\gamma^{0}}{2}iS^{+}(x;x') \!+\! \frac{1\!-\!{\rm e}^{i\phi\gamma^5}\gamma^{0}}{2}i\Tilde{S}^{+}(x;x')\right]
\,,\quad
\end{equation}
 where we made use of,
 ${\rm e}^{-\frac{i\phi}{2}\gamma^5}\left(i\gamma^{\!\mu}\nabla_{\mu} +|m|\right)
=\left(i\gamma^{\!\mu}\nabla_{\mu} +{\cal M}^\dagger\right)
       {\rm e}^{\frac{i\phi}{2}\gamma^5}$.
 This means that the split between the positive and negative frequency poles in~(\ref{3.11}) 
 is not  facilitated by the usual positive an negative energy projectors, 
 $P_\pm=\frac12(1\pm\gamma^0)$,
 but instead by more complicated projectors that involve chiral rotations,
 $P_\pm^5=\frac12(1\pm{\rm e}^{i\gamma^5\theta}\gamma^0)$.~\footnote{Notice that these operators are still proper projectors, in the sense that they obey, $[P_\pm^5]^2=P_\pm^5$
 and $P_+^5P_-^5=0=P_-^5P_+^5$, where we made use of 
 ${\rm e}^{i\gamma^5\phi}=\cos(\phi)+i\gamma^5\sin(\phi)$.}
This is a new result for the vacuum propagators of both 
 Dirac fermions and Majorana fermions on de Sitter space.

 When compared with the corresponding Dirac propagator on de Sitter, we see that -- in the special 
case when the mass is real -- our 
Majorana propagator~(\ref{3.12}--\ref{3.11}) becomes identical to  the Dirac 
propagator of Refs.~\cite{Candelas:1975du,Koksma:2009tc}, with the remark that our propagator represents 
generalization to the complex mass case. The fact that the two propagators agree 
should not surprise us in retrospect, because the propagators 
measure statistical properties of vacuum fluctuations, which -- in the absence of CP violation -- are identical  
for positive and negative frequency modes for Dirac particles, thus yielding statistically identical contributions 
from both frequency poles in the de Sitter vacuum. 
The same is true by construction~\footnote{Recall that the creation and annihilation operators are identical for both frequency poles.} for Majorana particles, {\it cf.} 
footnote~\ref{complex and real scalar}.

\subsection{Minkowski limit of the propagator}
\label{Minkowski limit of the propagator}

Here we consider the Minkowski limit of the 
propagator constructed in the previous section in Eq.~\eqref{3.11}. 
To do this, we will first expand the hypergeometric functions
 $iS^{+}(x;x')$ and $i\Tilde{S}^{+}(x;x')$  in Eq.~\eqref{3.11} using 
the identity~(9.131.2) from Ref.~\cite{Gradshteyn:2014},
\begin{equation}
    \label{3.13}
    \setlength{\jot}{15pt}
    \begin{split}
        iS^{+}(x;x') &= \frac{\Gamma\left(\frac{D}{2} + i\zeta\right)\Gamma\left(\frac{D-2}{2} - i\zeta\right)}{\left(4\pi\right)^{D/2}\Gamma(\frac{D}{2})}
\\        
        \times&\bigg\{\frac{\Gamma(\frac{D}{2})\Gamma\left(\frac{D}{2} - 1\right)}{\Gamma\left(\frac{D-2}{2} - i\zeta\right)\Gamma\left(\frac{D}{2}+i\zeta\right)}\left(\frac{y_{++}(x;x')}{4}\right)^{\frac{2-D}{2}}
\times{}_{2}F_{1}\left(1+i\zeta,-i\zeta;\frac{4-D}{2},\frac{y_{++}(x,x')}{4}\right) 
        \hspace{3cm}\\
 &+ \frac{\Gamma(\frac{D}{2})\Gamma\left(\frac{2-D}{2}\right)}{\Gamma\left(1+i\zeta\right)\Gamma\left(-i\zeta\right)}\times{}_{2}F_{1}\left(\frac{D-2}{2}-i\zeta,\frac{D}{2}+i\zeta;\frac{D}{2},\frac{y_{++}(x;x')}{4}\right)\bigg\}\hspace{-1cm}
    \end{split}
\end{equation}
\begin{eqnarray}
    \label{3.14}
    \setlength{\jot}{15pt}
        i\Tilde{S}^{+}(x;x') \!\!&=&\!\! \frac{\Gamma\left(\frac{D}{2} - i\zeta\right)\Gamma\left(\frac{D-2}{2} + i\zeta\right)}{\left(4\pi\right)^{D/2}\Gamma(\frac{D}{2})}
\\
\nonumber        
        &&\hskip -2cm
\times\,\bigg\{\frac{\Gamma(\frac{D}{2})\Gamma\left(\frac{D}{2} - 1\right)}{\Gamma\left(\frac{D-2}{2} + i\zeta\right)\Gamma\left(\frac{D}{2}-i\zeta\right)}\left(\frac{y_{++}(x,x')}{4}\right)^{\frac{2-D}{2}}
        \hskip -0.3cm
        \times{}_{2}F_{1}\left(1-i\zeta,+i\zeta;\frac{4-D}{2},\frac{y_{++}(x;x')}{4}\right)
        \hspace{0cm}
\\
&&\hskip -1cm
 +\, \frac{\Gamma(\frac{D}{2})\Gamma\left(\frac{2-D}{2}\right)}{\Gamma\left(1-i\zeta\right)\Gamma\left(i\zeta\right)}\times {}_{2}F_{1}\left(\frac{D-2}{2}+i\zeta,\frac{D}{2}-i\zeta;\frac{D}{2},\frac{y_{++}(x;x')}{4}\right)\bigg\}
 %\nonumber\\
% &&\hskip -2cm
%  =
\,,\quad
\end{eqnarray}
which transforms the propagator to the form that is convenient near the lightcone, where  
$y_{++}(x;x')\sim 0$.

The Minkowski limit is defined by $H\rightarrow 0$ and $a(\eta), a(\eta')\rightarrow 1$,
which also means that, when expanding the hypergeometric functions we can 
use that $(1\pm i\zeta)_n(\mp i\zeta)_n \simeq (|m|/H)^{2n}$ and 
$(y_{++}(x;x')/4)^n\simeq H^{2n}( \Delta x_{++}^{2})^n$,
where $\Delta x_{++}^{2}\equiv\Delta x_{++}^{2}(x;x')$ is defined in Eq.~(\ref{3.2}).
 Inserting these 
into~(\ref{3.13}--\ref{3.14}) yields (in the Minkowski limit), 
\begin{eqnarray}
	\label{ML.10}
    iS^{+}(x;x')\!\!&=&\!\! i\Tilde{S}^{+}(x;x') 
\\
\!\!&\simeq&\!\!
  \frac{\Gamma\left(\frac{D-2}{2}\right)\Gamma\left(\frac{4-D}{2}\right)\left(\frac{\lvert m \rvert}{H}\right)^{D-2}}{2\left(2\pi\right)^{D/2}}\;
  \frac{I_{\frac{D-2}{2}}\left(\lvert m \rvert\sqrt{\Delta x_{++}^{2}}\right)+I_{-\frac{D-2}{2}}\left(\lvert m \rvert\sqrt{\Delta x_{++}^{2}}\right)}{\left(\lvert m \rvert\sqrt{\Delta x_{++}^{2}}\right)^{ \frac{D-2}{2}}}
\,,\quad
\nonumber  
\end{eqnarray}
where $I_{\pm\nu}(z)$ denote modified Bessel functions of the first kind.
To get the last piece $\propto I_{-\frac{D-2}{2}}$ in Eq.~(\ref{ML.10}) 
we made use of the following asymptotic property of the gamma functions,
\begin{equation}
 \frac{\Gamma\left(\frac{D}{2} - i\zeta\right)\Gamma\left(\frac{D-2}{2} + i\zeta\right)}
     {\Gamma\left(1 - i\zeta\right)\Gamma\left( i\zeta\right)}
      \approx (i\zeta)^{D-2}\qquad ({\rm correct \; when }\; |i\zeta|\rightarrow \infty)
\,.\quad
\label{asymptotic gamma's}
\end{equation}
Now using the following definition,
\begin{equation}
	\label{ML.11}
    I_{\nu}(z) + I_{-\nu}(z) = \frac{2}{\Gamma\left(\nu\right)\Gamma\left(1-\nu\right)}K_{\nu}(z)
\,,
\end{equation}
where $K_{\nu}(z)$ denotes modified Bessel's function of the second kind,
one can recast Eq. (\ref{ML.10}) as,
\begin{equation}
	\label{ML.12}
    iS^{+}(x;x') = i\Tilde{S}^{+}(x;x') \simeq \frac{\left(\frac{\lvert m \rvert}{H}\right)^{D-2}}{\left(2\pi\right)^{D/2}}\;\frac{K_{\frac{D-2}{2}}\left(\lvert m \rvert\sqrt{\Delta x_{++}^{2}}\right)}{\left(\lvert m \rvert\sqrt{\Delta x_{++}^{2}}\right)^{\frac{D-2}{2}}}
\,.\quad
\end{equation}
From this it follows that the Minkowski limit the propagator~(\ref{3.11}) is,
\begin{equation}
	\label{ML.13}
    iS_{F}(x;x')\;\longrightarrow\; %\xrightarrow{\rm Mink.}
    \left[ iS_{F}(x;x')\right]_{\rm Mink} =
    \left(i\gamma^{\mu}\partial_{\mu} 
       + \mathcal{M}^{\dagger}\right)i\Delta_{\lvert m \rvert}(x;x')
\,,\quad
\end{equation}
where $i\Delta_{\lvert m \rvert}(x;x')$ denotes the massive scalar propagator in
Minkowski space (whose mass equals $|m|$), 
\begin{equation}
	\label{ML.14}
    i\Delta_{\lvert m \rvert}(x;x') =  \frac{\lvert m \rvert^{D-2}}{\left(2\pi\right)^{D/2}}\;
%\times
\frac{K_{\frac{D-2}{2}}\left(\lvert m \rvert\sqrt{\Delta x_{++}^{2}}\right)}{\left(\lvert m \rvert\sqrt{\Delta x_{++}^{2}}\right)^{\frac{D-2}{2}}}
\,.\quad
\end{equation}
The propagator~(\ref{ML.13}) obeys the Minkowski space Dirac equation
with the correct source,
 \begin{equation}
  \left(i\gamma^{\mu}\partial_{\mu} - {\cal M}\right) \left[ iS_{F}(x;x')\right]_{\rm Mink}
     = i\delta^D(x\!-\!x')
\,.\quad
 \label{EOM massive Majorana propagator: Mink}
\end{equation}
As it is evident, the chiral parts of the projectors from Eq.~\eqref{3.11} do not contribute in the Minkowski limit. It is important to ask whether the chiral part of the projector in the Majorana propagator in~\eqref{3.11} can contribute to CP-violation. 
This question can be properly addressed only when 
the propagator is used in loop calculations, in which vertices contain CP violation.
Since such calculations are quite involved, this question 
is beyond the scope of this paper.

\subsection{Charge conjugation}
\label{Charge conjugation}

Charge conjugation (CP) of the Majorana propagator is important for understanding leptogenesis which often involves CP-violating processes during the decay of massive Majorana neutrinos. To understand this better let us look at the CP transformation of the propagator given by,
\begin{equation}
    \label{3.17}
iS_{F}(x;x') \;\stackrel{\rm CP}{\longrightarrow}\; 
    \left(-i\gamma_{2}\right)%_{\alpha\lambda}
    \left(iS_{F}(x;x')\right)%_{\lambda\delta}
    ^{*}\left(-i\gamma_{2}\right)%
   % _{\delta\beta}  
   = - iS_{F}(x;x')
\,,
\end{equation}
which immediately follows from the Majorana conditions, 
\begin{equation}
\Psi(x)  = (-i\gamma^2)\Psi^*(x)\,,
   \qquad 
\bar\Psi(x)  = -\bar\Psi^*(x)(-i\gamma^2)
\label{Majorana condition: fields}
\end{equation}
and the propagator definition~(\ref{3.7b}).

To see how the property~(\ref{3.17}) pans out for our propagator~(\ref{3.11}), 
it is instructive to go through the mode decomposition~(\ref{2.18}---\ref{2.18b})
to understand how CP-violation realises~(\ref{3.17}). 
Let us begin our analysis with Eq.~(\ref{3.9}),
\begin{equation}
    \label{3.28}
    \begin{split}
       \left\langle T[\hat{\Tilde{\Psi}}_{M,\alpha}(x)\hat{\Tilde{\bar{\Psi}}}_{M,\beta}(x')]\right\rangle = \Theta(\eta - \eta')\sum_{h=\pm}\int \frac{d^{D-1}k}{(2\pi)^{D-1}}e^{i\vec{k}\cdot(\vec{x} - \vec{x}')}A_{h,\alpha}(\eta, \vec{k})\bar{A}_{h,\beta}(\eta', \vec{k}) \\- \Theta(\eta' - \eta)\sum_{h=\pm}\int\frac{d^{D-1}k}{(2\pi)^{D-1}}e^{-i\vec{k}\cdot(\vec{x} - \vec{x}')}A_{h,\alpha}(\eta, -\vec{k})\bar{A}_{h,\beta}(\eta', -\vec{k})
\,,
    \end{split}
\end{equation}
where  $A_{h,\alpha}(\eta, -\vec{k})$ and $\bar{A}_{h,\beta}(\eta', -\vec{k})$ 
are defined as,
\begin{eqnarray}
    \label{3.29}
    A_{h,\alpha}(\eta, -\vec{k}) \!\!&=&\!\! \begin{pmatrix}
    L_{h}(\eta, -\vec{k})\xi_{h}(-\vec{k})\\
     -he^{i\theta(\hat{k}_{x},\hat{k}_{y})}L^{*}_{h}(\eta, \vec{k})\xi_{h}(-\vec{k})
    \end{pmatrix}
\\
    \label{3.30}
    \bar{A}_{h,\beta}(\eta', -\vec{k}) \!\!&=&\!\!  \begin{pmatrix}
           -he^{-i\theta(\hat{k}_x,\hat{k}_{y})}L_{h}(\eta',\vec{k})\xi_{h}^{\dagger}(-\vec{k}) \;\;
                        & L^{*}_{h}(\eta', -\vec{k})\xi^{\dagger}_{h}(-\vec{k})
    \end{pmatrix}
\,.\quad
\end{eqnarray}
We can also see how $A_{h,\alpha}(\vec{k,\eta})\bar{A}_{h,\beta}(\eta', \vec{k})$ and $A_{h,\alpha}(-\vec{k,\eta})\bar{A}_{h,\beta}(\eta', -\vec{k})$
 transform to each other under charge conjugation,
\begin{equation}
    \label{3.31}
    (-i\gamma_{2})_{\alpha\lambda}\left(A_{h,\lambda}(\eta, \vec{k})\bar{A}_{h,\delta}(\eta', \vec{k})\right)^{*}(-i\gamma_{2})_{\delta\beta} = -A_{h,\alpha}(\eta, -\vec{k})\bar{A}_{h,\beta}(\eta', -\vec{k})
\end{equation}
\begin{equation}
    \label{3.32}
    (-i\gamma_{2})_{\alpha\lambda}\left(A_{h,\lambda}(\eta, -\vec{k})\bar{A}_{h,\delta}(\eta', -\vec{k})\right)^{*}(-i\gamma_{2})_{\delta\beta} = -A_{h,\alpha}(\eta, \vec{k})\bar{A}_{h,\beta}(\eta', \vec{k}).
\end{equation}
Applying the charge conjugation transformation to Eq.~\eqref{3.28} one obtains,
\begin{eqnarray}
    \label{3.33}
   &&\hskip -1.5cm    
    (-i\gamma_{2})_{\alpha\lambda}\Big(\Big\langle T\Big[\hat{\Tilde{\Psi}}_{M,\lambda}(x)
     \hat{\Tilde{\bar{\Psi}}}_{M,\delta}(x')\Big]\Big\rangle\Big)^{*}(-i\gamma_{2})_{\delta\beta} 
\\        
        \!\!&=&\!\!
         \Theta(\eta \!-\! \eta')\sum_{h = \pm}\int \frac{d^{D-1}k}{(2\pi)^{D-1}}e^{-i\vec{k}\cdot(\vec{x} - \vec{x}')}(-i\gamma_{2})_{\alpha\lambda}\left(A_{h,\lambda}(\eta, \vec{k})\bar{A}_{h,\delta}(\eta', \vec{k})\right)^{*}(-i\gamma_{2})_{\delta\beta}\hspace{.2cm}
\nonumber \\
  \!\!&&\!\! \hspace{0.cm}
  -\,\Theta(\eta' \!-\! \eta)\sum_{h=\pm}\int \frac{d^{D-1}k}{(2\pi)^{D-1}}e^{i\vec{k}\cdot(\vec{x}-\vec{x}')}(-i\gamma_{2})_{\alpha\lambda}\left(A_{h,\lambda}(\eta, -\vec{k})\bar{A}_{h,\delta}(\eta', -\vec{k})\right)^{*}(-i\gamma_{2})_{\delta\beta}
\hspace{.5cm}\;\nonumber\\
          \!\!&=&\!\!
-\Bigg[\Theta(\eta \!-\! \eta')\sum_{h = \pm}\int \frac{d^{D-1}k}{(2\pi)^{D-1}}
      e^{-i\vec{k}\cdot(\vec{x} - \vec{x}')} 
         A_{h,\alpha}(\eta, -\vec{k})\bar{A}_{h,\beta}(\eta', -\vec{k})
\hspace{.3cm}
 \nonumber
\\
          \!\!&&\!\!\hspace{1cm}
          - \,\Theta(\eta' \!-\! \eta)\sum_{h = \pm}\int \frac{d^{D-1}k}{(2\pi)^{D-1}}e^{i\vec{k}\cdot(\vec{x} - \vec{x}')}A_{h,\alpha}(\eta, \vec{k})\bar{A}_{h,\beta}(\eta', \vec{k})
    \Bigg].
\hspace{.2cm}
   \label{3.33b}
%   \end{split} 
\end{eqnarray}
If we were to calculate the integrals in Eq.~\eqref{3.33b}, we would obtain,
\begin{equation}
    \label{3.35}
    \setlength{\jot}{15pt}
    \begin{split}
        &
\left(a a'\right)^{-\frac{D-1}{2}}(-i\gamma_{2})_{\alpha\lambda}
 \Big(\Big\langle T\Big[\hat{\Tilde{\Psi}}_{M,\lambda}(x)\hat{\Tilde{\bar{\Psi}}}_{M,\delta}(x')
  \Big]\Big\rangle\Big)^{*}(-i\gamma_{2})_{\delta\beta} 
 \\       
&\hspace{0.5cm}
  =\  \bigg\{a\big(i\gamma^{\mu}\nabla_{\mu} + | m |\big)
\frac{H^{D-2}}{\sqrt{aa'}}
\bigg[\bigg(\frac{1 + \gamma^{0}}{2}\bigg)S^{-}(x;x') 
    + \bigg(\frac{1-\gamma^{0}}{2}\bigg)\Tilde{S}^{-}(x;x')\bigg]\bigg\}_{\alpha\beta}
\,,\hspace{0.1cm}
    \end{split}
\end{equation}
where $iS^{-}(x;x')$ and $i\Tilde{S}^{-}(x;x')$ in contrast to $iS^{+}(x;x')$ 
and $i\Tilde{S}^{+}(x;x')$ are given by,
\begin{equation}
    \label{3.36}
    \begin{split}
        iS^{-}(x;x') 
&= \frac{\Gamma\left(\frac{D}{2} \!+\! i\zeta\right)
              \Gamma\left(\frac{D-2}{2} \!-\! i\zeta\right)}
 {(4\pi)^{D/2}\Gamma(\frac{D}{2})}{}_{2}F_{1}
 \left(\frac{D}{2}\!+\!i\zeta,\frac{D\!-\!2}{2}\!-\!i\zeta;\frac{D}{2};1\!-\!\frac{y_{--}(x;x')}{4}\right)
 \\
        i\Tilde{S}^{-}(x;x') 
&= \frac{\Gamma\left(\frac{D}{2} \!-\! i\zeta\right)
               \Gamma\left(\frac{D-2}{2} \!+\! i\zeta\right)}
 {(4\pi)^{D/2}\Gamma(\frac{D}{2})}{}_{2}F_{1}
     \left(\frac{D}{2}\!-\!i\zeta,\frac{D\!-\!2}{2}\!+\!i\zeta;\frac{D}{2};1\!-\!\frac{y_{--}(x;x')}{4}\right)
.
    \end{split}
\end{equation}
Notice that, upon the charge conjugation transformation, we have the propagator 
in terms of $S^{-}(x;x')$ and $\Tilde{S}^{-}(x;x')$, which is obtained when the 
time ordering for the Dyson propagator is imposed. In order to understand this result,
recall the St\"{u}ckelberg rule, which states that propagation of particles in the positive time direction (governed by the time ordering in the Feynman propagator) is identical 
to propagation of antiparticles in the negative time direction. Reversing the direction of time,
the rule tells us that propagation of particles in the negative time direction
is equivalent to propagation of antiparticles in the positive time direction, which explains
the Dyson time order in Eq.~(\ref{3.36}). Since we are dealing with
 Majorana fermions, propagation of antiparticles in the positive time direction must be 
 related in a simple way to propagation of particles in the positive time direction. 
 To see that, let us have a look at Eq.~(\ref{3.33b}), in which 
 we see that the momentum integrals contain opposite phases when compared with that in
Eq.~(\ref{3.9}). This is cured by flipping the sign of the integrated momenta 
resulting in, 
\begin{equation}
    \label{3.37}
    \begin{split}
      &  \hskip -1cm
(-i\gamma_{2})_{\alpha\lambda}\Big(\Big\langle T\Big[\hat{\Tilde{\Psi}}_{M,\lambda}(x)\hat{\Tilde{\bar{\Psi}}}_{M,\delta}(x')\Big]\Big\rangle\Big)^{*}(-i\gamma_{2})_{\delta\beta}
\\
 &= -\bigg[\Theta(\eta - \eta')\sum_{h = \pm}\int \frac{d^{D-1}k}{(2\pi)^{D-1}}e^{i\vec{k}\cdot(\vec{x} - \vec{x}')} A_{h,\alpha}(\eta, \vec{k})\bar{A}_{h,\beta}(\eta', \vec{k}) 
% \nonumber
\\
 &  \hskip 1cm
 -\, \Theta(\eta' - \eta)\sum_{h = \pm}\int \frac{d^{D-1}k}{(2\pi)^{D-1}}e^{-i\vec{k}\cdot(\vec{x} - \vec{x}')}A_{h,\alpha}(\eta, -\vec{k})\bar{A}_{h,\beta}(\eta', -\vec{k})\bigg].
% \nonumber
    \end{split}\quad
\end{equation}
Evaluating this integral we get the result that we calculated before in Eq. \eqref{3.11}, however with an overall negative sign, as expected,
\begin{equation}
    \label{3.38}
    \setlength{\jot}{15pt}
    \begin{split}
     &\hskip -1cm
(-i\gamma_{2})_{\alpha\lambda}\Big(\Big\langle T\Big[\hat{\Tilde{\Psi}}_{M,\lambda}(x)\hat{\Tilde{\bar{\Psi}}}_{M,\delta}(x')\Big]\Big\rangle\Big)^{*}(-i\gamma_{2})_{\delta\beta}
\\
&=  -\bigg\{a\big(i\gamma^{\mu}\nabla_{\mu} + | m |\big)\frac{H^{D-2}}{\sqrt{aa'}}\left[\frac{1 + \gamma^{0}}{2}S^{+}(x;x') + \frac{1-\gamma^{0}}{2}\Tilde{S}^{+}(x;x')\right]\bigg\}_{\alpha\beta}
\\
& = -\Big\langle T\Big[\hat{\Tilde{\Psi}}_{M}(x)\hat{\Tilde{\bar{\Psi}}}_{M}(x')\Big]\Big\rangle
       \,.\hspace{.5cm}
    \end{split}
\end{equation}
Now our result is in accordance with our expectation in Eq.~\eqref{3.17}, in which  
it is expressed in terms of time ordered 
$S^{+}(x;x')$ and $\Tilde{S}^{+}(x;x')$, which compose
the Feynman propagator~(\ref{chiral rotation}--\ref{3.11}).

\begin{figure}[h]
\vskip -.2cm
    %\centering
 {{\includegraphics[width=0.4\textwidth]{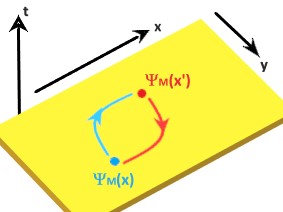} }}%
    \qquad\qquad
{\includegraphics[width=0.4\textwidth]{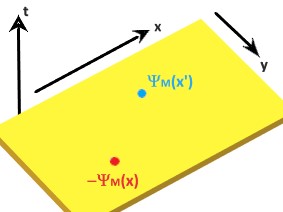}}
\vskip -.cm
   \caption{
   {\it Left panel} shows that the two Majorana fermions $\Psi(x)$ and $\Psi(x')$ confined in 
   two spatial dimensions at positions $x$ and $x'$, denoted by blue and red dots, respectively. 
The arrows indicated how they are exchanged (half-braided) with each other. 
{\it Right panel} shows that, after the exchange: $\Psi(x)\rightarrow\Psi(x')$ and $\Psi(x')\rightarrow-\Psi(x)$. Although the exchange can be visualised by how the blue and red dots have changed their position, this is only for a visual representation. However, in reality such a 
distinction is not possible.
    }
\label{fig:Fig3} 
\end{figure}

An important question is whether the minus sign~(\ref{3.38}) 
can have any physical significance. Here we refrain from making any detailed analysis, but just
remark that fermionic signs can lead to observable effect in two dimensional 
($D=1+2=3$) topological systems, a notable example being fermionic systems 
considered for building quantum computers in which braiding of the fermionic wave 
function plays an important role~\cite{Alicea:2010},\cite{Ivanov:2000mjr}.  The charge conjugation operation of the Majorana propagator can be seen as the half-braiding~\footnote{The operation performed here is an exchange and thus it corresponds to a half-braid. A full braid involves the majorana fermion going around the other fermion, back to its original  site.} operation between two Majorana fermions in two spatial dimensions, such that the transformation $\mathcal{C}\hat{\Tilde{\Psi}}_{M}(x)$ and $\hat{\Tilde{\bar{\Psi}}}_{M}(x')\mathcal{C}^{\dagger}$ is similar to exchange of particles as shown in Fig.~\ref{fig:Fig3} (left panel). After this half-braiding transformation, the particles exchange their position, however, one of the Majorana fermion operator incurs a negative sign which can be seen in Fig.~\ref{fig:Fig3} (right panel). Therefore the half-braiding operation follows the transformation rule $\hat{\Tilde{\Psi}}_{M}(x)\rightarrow\hat{\Tilde{\Psi}}_{M}(x')$ and  $\hat{\Tilde{\Psi}}_{M}(x')\rightarrow-\hat{\Tilde{\Psi}}_{M}(x)$. It must be noted that the negative sign that we see in the half-braiding operation in two dimensions is physically realized as different qubit states that are topologically protected~\cite{Beenakker:2020}. The question whether the charge conjugation operation of Majorana fermions in general $D$ dimensions has any physical relevance, and  
whether it can be related 
to the braiding operation in $2+1$ dimensions requires further analysis.

%%%%%%%%%%%%%%%%%%%%%%%%%%%%%
%%%    O N E  L O O P  E F F E C T I V E  A C T I O N    %%%
%%%%%%%%%%%%%%%%%%%%%%%%%%%%%

\section{One-Loop Effective Action}
\label{One-Loop Effective Action}
\label{Section-4}

The one-loop effective action for Majorana fermions is given 
by~\footnote{When compared with  the
Dirac fermions, the one-loop effective action for the Majorana fermions~(\ref{4.1}) contains 
a factor $1/2$ in front of the trace, which can be understood by noting that 
$\bar\psi_M$ and $\psi_M$ are dependent variables. Recall that analogous difference occurs
between effective actions for real and complex scalar fields, as the effective action
for a complex scalar field is equivalent to that of two real scalars.} 
\begin{equation}
    \label{4.1}
    \Gamma^{(1)}_{\text{Maj}} = -\frac{i}{2} 
\text{Tr}\left[\text{log}\left(\sqrt{-g}\left(i\gamma^{\mu}\nabla_{\mu} 
 - \Tilde{\mathcal{M}}\right)\right)\right]
  =-\frac{i}{2} \text{Tr}\left[\text{log}\left(\sqrt{-g}\left(i\gamma^{\mu}\nabla_{\mu} 
 -|m|\right)\right)\right]
 \,,\quad
\end{equation}
where the second equality follows from Eq.~(\ref{chiral rotation}) and
the observation that ${\rm Tr}[\gamma^5]=0$.
%{chiral rotation}{3.7a}
Taking a derivative of~(\ref{4.1}) with respect to $|m|$ and integrating over 
$|m|$ gives an equvalent expression (up to an irrelevant integration constant),
\begin{equation}
    \label{4.2}
    \Gamma^{(1)}_{\text{Maj}} = \frac{1}{2}\int^{|m|} {\rm d}\Tilde m
     \text{Tr}\left[\sqrt{-g}iS^{ab}_{F}(x;x')\right].
\end{equation}
where $\tilde m>0$ is a real mass parameter.
Since the integral in~(\ref{4.2}) is idential to that which occurs for the Dirac fermions,
we can use Eq.~(93) from Ref.~\cite{Koksma:2009tc},
\begin{equation}
    \label{4.10}
    \Gamma^{(1)}_{\text{Maj}} 
    = \frac{1}{2}\int {\rm d}^{D}x\sqrt{-g}\frac{H^{D-2}}{(2\pi)^{D/2}}
    \left\{\Gamma\left(1 \!-\! \frac{D}{2}\right)\int^{ \rvert m \lvert }{\rm d}\Tilde m 
     \frac{\Tilde{m}\Gamma\left(\frac{D}{2}\!+\!i\Tilde{\zeta}\right)
          \Gamma\left(\frac{D}{2} \!-\! i\Tilde{\zeta}\right)}
          {\Gamma\left(1\!+\! i\Tilde{\zeta}\right)
           \Gamma\left(1\!-\! i\Tilde{\zeta}\right)}\right\}
,\quad
\end{equation}
where $\Tilde \zeta = \Tilde m/H$.
This action can be expanded around $D=4$, and its convenient rewriting is, 
\begin{eqnarray}
    \label{4.10b}
    \Gamma^{(1)}_{\text{Maj}} \!\!&=&\!\! \int {\rm d}^{D}x\sqrt{-g}\bigg\{
    \frac{\mu^{D-4}}{2(2\pi)^{D/2}}
    \Gamma\left(1 \!-\! \frac{D}{2}\right)\int^{ \rvert m \lvert }{\rm d}\Tilde m \Tilde{m}
     \bigg[H^2\!+\!\Tilde m^2
\nonumber\\     
    \!\!&&\!\! \hskip 2.6cm +\,\frac{D\!-\!4}{2}H^2
     \left(\psi\left(1\!+\! i\Tilde{\zeta}\right)
           +\psi\left(1\!-\! i\Tilde{\zeta}\right)+ 2 + \ln\Big(\frac{H^2}{\mu^2}\Big)
     \right)\bigg]\bigg\}
\,,\qquad
\end{eqnarray}
where we made use of 
$H^{D-2}=\mu^{D-4} H^2\Big[1+\frac{D-4}{2}\ln\big(H^2/\mu^2\big)
    +{\cal O}\big((D-4)^2\big)\Big]$.
Now, keeping in mind that, 
\begin{equation}
\Gamma\Big(1-\frac{D}{2}\Big)= \frac{2}{D-4} + \gamma_E -1 
\,,
\label{effective action: isolating div}
\end{equation}
where $ \gamma_E=\psi(1)\approx 0.57\cdots$ is the Euler-Mascheroni constant, one sees
that the divergences in~(\ref{4.10b})
 can be removed by adding the counterterm action of the form, 
\begin{equation}
S_{\rm ct} = \int {\rm d}^{D}x\sqrt{-g}\bigg[ \alpha \frac{R}{D(D-1)} + \beta\bigg]
\,,\qquad R=D(D-1) H^2
\,,\qquad
\label{counterterm action}
\end{equation}
where 
\begin{eqnarray}
 \alpha  \!\!&=&\!\! -\frac{\mu^{D-4} }{4(2\pi)^{D/2}}  
    \Gamma\left(1 \!-\! \frac{D}{2}\right)m^2 + \alpha_{\rm f}
\,,\qquad \nonumber\\
 \beta  \!\!&=&\!\! -\frac{\mu^{D-4} }{8(2\pi)^{D/2}}  
       \Gamma\left(1 \!-\! \frac{D}{2}\right) m^4 + \beta_{\rm f}
\,,\qquad
\label{counterterm action 2}
\end{eqnarray}
resulting in the following renormalized one-loop effective action, 
\begin{eqnarray}
    \Gamma^{(1)\rm ren}_{\text{Maj}} \!\!&=&\!\! \int {\rm d}^{4}x\sqrt{-g}\bigg\{
    \frac{H^2}{8\pi^2}
     \bigg[ \int^{ \rvert m \lvert }{\rm d}\Tilde m \Tilde{m}
     \left(\psi\left(1\!+\! i\Tilde{\zeta}\right)
           \!+\!\psi\left(1\!-\! i\Tilde{\zeta}\right)
     \right)\!+\! m^2 \!+\! \frac{m^2}{2}\ln\bigg(\frac{H^2}{\mu^2}\bigg)\bigg]\bigg\}
\qquad
\nonumber\\
 \!\!&+&\!\!\hskip 0cm S_{\rm HE}
\,,\qquad
\label{renormalized action}
\end{eqnarray}
where $S_{\rm HE}$ is the Hilbert-Einstein action with a renormalized Newton constant 
and cosmological constant, whereas the finite parts of the counterterm couplings 
$\alpha_{\rm f}$ and $\beta_{\rm f}$ from~(\ref{counterterm action 2}) 
were absorbed in the (renormalized)
Newton constant and cosmological constant, respectively.
The one-loop effective action~(\ref{renormalized action}) equals $1/2$ of the corresponding
Dirac field one-loop effective action, {\it cf.} Ref.~\cite{Koksma:2009tc}. This factor $1/2$ can 
be attributed to the reduced number of degrees of freedom carried by the Majorana particle.
In the above renormalization we used a non-minimal subtraction scheme
and assumed that the Majorana mass is a parameter. 
If the mass is generated by a scalar field condensate,
which is what was assumed in Ref.~\cite{Koksma:2009tc}, a different (more complicated) counterterm action 
is needed. For details of the renormalization procedure in that case 
we refer to Ref.~\cite{Koksma:2009tc}.

%%%%%%%%%%%%%%%%%%%%%%%%%
%%%    M U L T I F L A V O R  F E R M I O N S  %%%
%%%         A N D   C P   V I O L A T I O N         %%%
%%%%%%%%%%%%%%%%%%%%%%%%%

\section{Multiflavor fermions and CP violation}
\label{Multiflavor fermions and CP violation}
\label{Section-5}

After analyzing the propagator for a single Majorana fermion
on de Sitter space, here we extend the analysis to many mixing Majorana particles, which can  
serve as a starting point for a better understanding of CP violation.

\subsection{Multiflavor Majorana Fermions}
\label{Multiflavor Majorana Fermions}

The Lagrangian for the multiflavor Majorana fermion, $\Lambda{_{I}}(x)$, $(I=1,2,\cdots, n)$, 
where $n$ is the number of flavors, is the following natural generalization of Eq.~(\ref{2.1M}),
\begin{equation}
    \label{5.1}
    \mathcal{L} = \frac14 i\bar{\Lambda}_{_{I}}(x)\gamma^{\mu}
    \!\stackrel{\leftrightarrow}{\nabla}_{\mu}\!
\Lambda_{_{I}}(x) 
      - \frac12\bar{\Lambda}_{_{I}}(x)\mathcal{M}_{_{IJ}}\Lambda_{_{J}}(x)
      + {\cal L}_{\rm int}
\,,
\end{equation}
where $\mathcal{M}_{_{IJ}}$ is a complex symmetric Majorana mass matrix,
\begin{equation}
    \label{mass matrix: def}
   \pmb{\mathcal{M}}
    = \begin{pmatrix}
    \mathbf{m}^{\dagger} & 0\\
    0 & \mathbf{m}
    \end{pmatrix}
\,,\qquad 
\end{equation}
where $\mathbf{m}$ is a symmetric complex scalar 
(non-spinorial) $n\times n$ mass matrix,
and ${\cal L}_{\rm int}$ is an interaction Lagrangian, which in the simple case
of Yukawa couplings can be written as, 
\begin{equation}
{\cal L}_{\rm int}=\bar{\Lambda}_{_{I}}(x)\phi(x)\mathcal{Y}_{_{IJ}}\Lambda_{_{J}}(x)
\,,
\label{Yukawa coupling matrix}
\end{equation}
where $\mathcal{Y}_{_{IJ}}$ is the symmetric complex Yukawa matrix of couplings,
\begin{equation}
    \label{yukawa matrix: def}
   \pmb{\mathcal{Y}}
    = \begin{pmatrix}
    \mathbf{y}^{\dagger} & 0\\
    0 & \mathbf{y}
    \end{pmatrix}
\,,\qquad 
\end{equation}
where $\mathbf{y}$ is a symmetric scalar complex matrix of couplings,
and $\phi(x)$ is a scalar field condensate (which in general situations could be a matrix).
Next, after rescaling the Majorana fields as, 
$\Lambda_{_{I}}(x)\rightarrow a^{\frac{D-1}{2}}\Lambda_{_{I}}(x)=\Tilde{\Lambda}_{_{I}}(x)$. The Majorana field obeys the Majorana condition~,
\begin{equation}
    \label{5.2}
    \Tilde{\Lambda}_{_{I}}(x) = \Tilde{\Lambda}^{c}_{_{I}}(x) = -i\gamma^{2}\Tilde{\Lambda}^{*}_{_{I}}(x),
\end{equation}
from which we can define $\Tilde{\Lambda}_{_{I}}(x)$ as,
\begin{equation}
\label{5.3}
    \Tilde{\Lambda}_{_{I}}(x) = \begin{pmatrix}
    \lambda_{_{I}}(x)\\
    \epsilon\lambda^{*}_{_{I}}(x)
    \end{pmatrix},
\end{equation}
and the Dirac equation that results from  Eq.\eqref{5.1} is given by,
\begin{equation}
    \label{5.4}
    \left(i\gamma^{b}\partial_{b} - a\mathcal{M}\right)_{_{IJ}}\Tilde{\Lambda}_{_{J}}(x) = 0
    \,,
\end{equation}
where we neglected the Yukawa term~(\ref{Yukawa coupling matrix}).
Here the Majorana field $\Tilde{\Lambda}_{_{J}}(x)$ is in the flavor mixing basis and we can write it in the mass diagonal basis as follows,
\begin{equation}
    \label{5.5}
    \Tilde{\Lambda}_{_{J}}(x) = V_{_{JK}}^{\dagger}\Lambda_{_{K}}^{(d)}(x) 
     = \left[\mathbf{V}^{\dagger}\cdot\pmb{\Lambda}^{(d)}(x)\right]_{J},
\end{equation}
and
\begin{equation}
    \label{5.6}
    \bar{\Tilde{\Lambda}}_{_{J}}(x) = \bar{\Tilde{\Lambda}}^{(d)}_{_{K}}(x)U_{_{KJ}} 
    = \left[\bar{\pmb{\Lambda}}^{(d)}(x)\cdot\mathbf{U}\right]_J
\,,
\end{equation}
where unitary matrices $\mathbf{V}$ and $\mathbf{U}$ are defined as follows,
\begin{equation}
    \label{5.7}
    \mathbf{V} = \begin{pmatrix}
    \mathbf{U}_{_{L}} & 0\\
    0 & \mathbf{U}_{_{R}}
    \end{pmatrix}
\,,\qquad
 %   \label{5.8}
    \mathbf{U} = \begin{pmatrix}
    \mathbf{U}_{_{R}} & 0\\
    0 & \mathbf{U}_{_{L}}
    \end{pmatrix}
\,.\qquad
\end{equation}
These matrices are used for the bi-unitary diagonalization of the flavor mass matrix in the case of Dirac fermions. However in the case of Majorana fermions, due to the Majorana condition  
the mass matrix is symmetric and therefore we do not need a general bi-unitary transformation. 
In order to understand the implication of this fact, note first that both $\pmb{\Lambda}$ and
$\pmb{\Lambda}^{(d)}$ should obey the Majorana condition, 
 $\mathbf{V}^{\dagger}\cdot\mathbf{\Lambda} 
= -i\gamma^{2}\mathbf{V}^{T}\cdot\mathbf{\Lambda}^{*}$,
which translates to $\mathbf{U}_{_{R}} = \mathbf{U}^{*}_{_{L}}$, such that 
Eq.~(\ref{5.7}) simplifies to,
\begin{equation}
    \label{5.9}
    \begin{split}
        \mathbf{V}\rightarrow\mathbf{X} = \begin{pmatrix}
        \mathbf{U}_{_{L}} & 0\\
        0 & \mathbf{U}^{*}_{_{L}}
        \end{pmatrix}
\,,\qquad
        \mathbf{U}\rightarrow\mathbf{Y} = \begin{pmatrix}
        \mathbf{U}^{*}_{_{L}} & 0\\
        0 & \mathbf{U}_{_{L}}
        \end{pmatrix}
\,.
    \end{split}
\qquad
\end{equation}
Eq.~\eqref{5.4} can be diagnonalized\footnote{The diagonalization of the Majorana mass matrix 
uses the Takagi diagonalization, for which the mass matrix is complex
and symmetric and the resulting diagonalized mass matrix is real and non-negative.
Some details of the diagonalization procedure can be found 
in Appendix~E, see in particular Eqs.~(\ref{D.3}--\ref{D.4}).}
by writing it as follows,
\begin{equation}
    \label{5.10}
    \left(i\gamma^{b}\partial_{b}\mathbf{1}-a\pmb{\mathcal{M}}\right)
    \cdot\mathbf{X}^{\dagger}\cdot\pmb{\Lambda}^{(d)}(x) = 0
\,,
\end{equation}
then we can multiply this equation by $\mathbf{Y}$ from the left which gives us the following result,
\begin{equation}
    \label{5.11}
    i\mathbf{Y}\cdot\gamma^{b}\partial_{b}\mathbf{X}^{\dagger}\cdot\pmb{\Lambda}^{(d)}(x) 
     - a\mathbf{Y}\cdot\pmb{\mathcal{M}}\cdot\mathbf{X}^{\dagger}\cdot\pmb{\Lambda}^{(d)}(x)
       = 0
\,.
\end{equation}
The term $\mathbf{Y}\cdot\gamma^{b}\partial_{b}\mathbf{X}^{\dagger} $ 
can be reduced as follows,
\begin{equation}
    \label{5.12}
    \begin{split}
        \mathbf{Y}\cdot\gamma^{b}\partial_{b}\mathbf{X}^{\dagger} 
        &= \begin{pmatrix}
    \mathbf{U}^{*}_{_{L}} & 0\\
    0 & \mathbf{U}_{_{L}}
    \end{pmatrix}\cdot\begin{pmatrix}
    0 & \sigma^{\mu}\partial_{\mu}\\
    \bar{\sigma}^{\mu}\partial_{\mu} & 0
    \end{pmatrix}\cdot\begin{pmatrix}
    \mathbf{U}^{\dagger}_{_{L}} & 0\\
    0 & \mathbf{U}^{T}_{_{L}}
    \end{pmatrix}\\
     &= \begin{pmatrix}
         0 & \sigma^{\mu}\partial_{\mu}\\
         \bar{\sigma}^{\mu}\partial_{\mu} & 0
     \end{pmatrix}
%\hspace{4cm}
%\\
     = \gamma^{b}\partial_{b}\mathbf{1}
\,,\hspace{5.5cm}
    \end{split}
\end{equation}
and thus Eq.~\eqref{5.11} can be written as,
\begin{equation}
    \label{5.13}
    \left(i\gamma^{b}\partial_{b}\mathbf{1} - a|\pmb{\mathcal{M}}^{(d)}|\right)
        \cdot\pmb{\Lambda}^{(d)}(x) = 0
\,,
\end{equation}
where $|\pmb{\mathcal{M}}^{(d)}|$ is the diagonalized mass matrix, which is real and non-negative. The diagonalization of the mass matrix from Eq.~\eqref{5.11} is given in Appendix~E, 
from which we have made use of the result in Eq.~\eqref{D.2}. 
We can then express the multiflavor Majorana fermion field operator 
in the mass diagonal basis as follows,
\begin{equation}
    \label{5.14}
    \hat{\pmb{\Lambda}}^{(d)}(x) = \sum_{h=\pm}\int\frac{d^{D-1}k}{(2\pi)^{D-1}}e^{i\vec{k}\cdot\vec{x}}\mathbf{A}^{(d)}_{h}(\eta, \vec{k})
     \cdot\left[\hat{\mathbf{b}}_{h}(\vec{k}) 
           + \hat{\mathbf{b}}^{\dagger}_{h}(-\vec{k}) \right],
\end{equation}
where $\hat{\mathbf{b}}_{h}(\vec{k}) $ and $\hat{\mathbf{b}}^{\dagger}_{h}(\vec{k})$
are the annihilation and creation operators, which are vectors in the flavor space, 
$\mathbf{A}^{(d)}_{h}(\eta, \vec{k})$ is the mode function matrix, which is 
diagonal in the basis in which the mass is diagonal, and it is defined by,
\begin{equation}
    \label{5.15}
    \begin{split}
    \mathbf{A}^{(d)}_{h}(\eta, \vec{k}) = \begin{pmatrix}
    \mathbf{L}^{(d)}_{h}(\eta, \vec{k})\cdot\pmb{\xi}_{h}(\vec{k})\\
    h\mathbf{L}^{*(d)}_{h}(\eta, -\vec{k})\cdot\pmb{\xi}^{*}_{h}(-\vec{k})
    \end{pmatrix}
    = \begin{pmatrix}
    \mathbf{L}^{(d)}_{h}(\eta, \vec{k})\cdot\pmb{\xi}_{h}(\vec{k})\\
    e^{i\phi'}\mathbf{L}^{*(d)}_{h}(\eta, -\vec{k})\cdot\pmb{\xi}_{h}(\vec{k})
    \end{pmatrix}
\,,\quad
    \end{split}
\end{equation}
where $\pmb{\xi}_{h}(\vec{k})$ builds an $n-$component flavor vector,
\begin{equation}
    \label{5.16}
    [\pmb{\xi}_{h}(\vec{k})]^T = \big[\xi_{h}(\vec{k}),\xi_{h}(\vec{k}),\cdots
            \big]_{_{n}},
\end{equation}
and similarly $\mathbf{L}^{(d)}_{h}(\eta, \vec{k})$ is the mode function 
$(n\times n)$ matrix in the diagonal basis,
\begin{equation}
    \label{5.17}
    \mathbf{L}^{(d)}_{h}(\eta, \vec{k}) = \text{diag}\left[L^{(d)}_{_{h,11}},L^{(d)}_{_{h,11}},....,L^{(d)}_{_{h,nn}}\right]
\,.\quad
\end{equation}
In Eq. \eqref{5.15} $\phi' = \frac{\pi(h-1)}{2} + \theta(\hat{k}_{x},\hat{k}_{x})$,
and it is identical
to the phase in Eq.~\eqref{2.31}, making use of the property in~\eqref{A.5} and the fact that helicity $h$ can be written as $h=e^{i\frac{\pi(h-1)}{2}}$. 
Furthermore, the solutions of the Dirac equation in the mass diagonal basis 
in Eq. \eqref{5.13} can be found in the same way as we have done for the single flavor case in 
section~\ref{Section-3},
\begin{eqnarray}
    \label{5.18}
    {\rm e}^{-\frac{i\phi'}{2}}\mathbf{L}^{(d)}_{h}(\eta, \vec{k}) 
    \!\!&=&\!\! \frac{1}{2}\sqrt{-\frac{\pi\|\textbf{k}\|\eta}{4}}
    \left[{\rm e}^{\frac{i\pi\pmb\nu^{\dagger}}{2}}
          \cdot\pmb{\mathcal{H}}^{(1)}_{\pmb\nu^{\dagger}}(-\|\textbf{k}\|\eta) 
     + h {\rm e}^{\frac{i\pi\pmb\nu}{2}}\cdot\pmb{\mathcal{H}}^{(1)}_{\pmb\nu}(-\|\textbf{k}\|\eta)
     \right]
\\
    \label{5.19}
    {\rm e}^{\frac{i\phi'}{2}}\mathbf{L}^{*(d)}_{h}(\eta, -\vec{k}) 
      \!\!&=&\!\! \frac{1}{2}\sqrt{-\frac{\pi\|\textbf{k}\|\eta}{4}}
     \left[{\rm e}^{\frac{i\pi\pmb\nu^{\dagger}}{2}}
     \cdot\pmb{\mathcal{H}}^{(1)}_{\pmb\nu^{\dagger}}(-\|\textbf{k}\|\eta) 
     - h {\rm e}^{\frac{i\pi\pmb\nu}{2}}\cdot\pmb{\mathcal{H}}^{(1)}_{\pmb\nu}(-\|\textbf{k}\|\eta)
     \right]
\\
    \label{5.20}
    {\rm e}^{\frac{i\phi'}{2}}\mathbf{L}^{*(d)}_{h}(\eta, \vec{k}) 
     \!\!&=&\!\! \frac{1}{2}\sqrt{-\frac{\pi\|\textbf{k}\|\eta}{4}}\left[
    {\rm e}^{\frac{-i\pi\pmb\nu}{2}}\cdot\pmb{\mathcal{H}}^{(2)}_{\pmb\nu}(-\|\textbf{k}\|\eta) 
    + h {\rm e}^{\frac{-i\pi\pmb\nu^{\dagger}}{2}}
           \cdot\pmb{\mathcal{H}}^{(2)}_{\pmb\nu^{\dagger}}(-\|\textbf{k}\|\eta)\right]
\\
    \label{5.21}
    {\rm e}^{\frac{-i\phi'}{2}}\mathbf{L}^{(d)}_{h}(\eta, -\vec{k}) 
     \!\!&=&\!\! \frac{1}{2}\sqrt{-\frac{\pi\|\textbf{k}\|\eta}{4}}
    \left[{\rm e}^{\frac{-i\pi\pmb\nu}{2}}\cdot\pmb{\mathcal{H}}^{(2)}_{\pmb\nu}(-\|\textbf{k}\|\eta) 
    - h {\rm e}^{\frac{-i\pi\pmb\nu^{\dagger}}{2}}
          \cdot\pmb{\mathcal{H}}^{(2)}_{\pmb\nu^{\dagger}}(-\|\textbf{k}\|\eta)\right]
,\qquad
\end{eqnarray}
where 
${\rm e}^{\frac{i\pi\pmb\nu}{2}}\cdot\pmb{\mathcal{H}}^{(1),(2)}_{\pmb\nu}(-\|\textbf{k}\|\eta)$
 and ${\rm e}^{\frac{i\pi\pmb\nu^{\dagger}}{2}}
 \cdot\pmb{\mathcal{H}}^{(1),(2)}_{\pmb\nu^{\dagger}}(-\|\textbf{k}\|\eta)$ 
 denote diagonal Hankel function matrices, 
% $\mathcal{H}}^{(1)}_{\pmb\nu^{\dagger}}(z)$ and 
%$\mathcal{H}}^{(2)}_{\pmb\nu^{\dagger}}(z)$ 
%
\begin{eqnarray}
    \label{B.12}
    {\rm e}^{\frac{i\pi\pmb\nu}{2}}
    \!\cdot\pmb{\mathcal{H}}^{(1),(2)}_{\nu}(-\|\textbf{k}\|\eta) 
    \!\!&=&\!\! \begin{pmatrix}
    \!{\rm e}^{\frac{i\pi\nu_{11}}{2}} & & \\
    & \ddots & \\
    & & {\rm e}^{\frac{i\pi\nu_{nn}}{2}}
    \end{pmatrix}
    \begin{pmatrix}
    \!\mathcal{H}^{(1),(2)}_{\nu_{11}}(-\|\textbf{k}\|\eta) & & \\
    & \ddots & \\
    & & \mathcal{H}^{(1),(2)}_{\nu_{nn}}(-\|\textbf{k}\|\eta)
    \end{pmatrix}
\\
    \label{B.13}
    {\rm e}^{\frac{i\pi\pmb\nu^{\dagger}}{2}}\!
    \cdot\pmb{\mathcal{H}}^{(1),(2)}_{\pmb\nu^{\dagger}}(-\|\textbf{k}\|\eta) 
    \!\!&=&\!\! \begin{pmatrix}
    \!{\rm e}^{\frac{i\pi\nu^{\dagger}_{11}}{2}} & & \\
    & \ddots & \\
    & & {\rm e}^{\frac{i\pi\nu^{\dagger}_{nn}}{2}}
    \end{pmatrix}
    \begin{pmatrix}
    \!\mathcal{H}^{(1),(2)}_{\nu^{\dagger}_{11}}(-\|\textbf{k}\|\eta) & & \\
    & \ddots & \\
    & & \mathcal{H}^{(1),(2)}_{\nu^{\dagger}_{nn}}(-\|\textbf{k}\|\eta)
    \end{pmatrix}\!.
\qquad\;\;
\end{eqnarray}
where ({\it cf.} Eq.~(\ref{D.3})),
\begin{equation}
    \label{5.22A}
        \pmb{\nu} = \frac{1}{2} + i  \pmb{\zeta}^{(d)} 
\,,\qquad
        \pmb{\nu}^{\dagger} = \frac{1}{2} - i\pmb{\zeta}^{(d)}
\,,\qquad        \pmb{\zeta}^{(d)} = \frac{\lvert \pmb{m}^{(d)} \rvert}{H}
\,
\end{equation}
are $n\times n$ diagonal matrices in flavor space.

\subsection{Multiflavor Majorana propagator}
\label{Multiflavor Majorana propagator}

The multiflavor Majorana propagator obeys the equation of motion,
\begin{equation}
    \label{5.22}
    \left(i\gamma^{b}\partial_{b} - a\mathcal{M}\right)_{_{IP}}iS_{_{PJ}}(x;x')
       = i\delta_{IJ}\delta^{D}(x\!-\!x'),
\end{equation}
where $\delta_{IJ}$ denotes the Kronecker delta in flavor space and
the propagator $iS_{_{PJ}}(x;x')$ is defined as (see Eq.~(\ref{3.7b})),
\begin{equation}
    \label{5.23}
    \begin{split}
        iS_{_{PJ}}(x;x') &=  \left[i\mathbf{S}_F(x;x')\right]_{PJ}
\,,\quad
\\
 i\mathbf{S}_F(x;x') 
  &= \Big\langle \text{T}\Big[\hat{\Tilde{\pmb{\Lambda}}}(x)\hat{\Tilde{\bar{\pmb{\Lambda}}}}(x')
       \Big]\Big\rangle 
= \Theta(\eta \!-\! \eta')\Big\langle \hat{\Tilde{\pmb{\Lambda}}}(x) 
         \hat{\Tilde{\bar{\pmb{\Lambda}}}}(x')\Big\rangle 
- \Theta(\eta' \!-\! \eta)\Big\langle \hat{\Tilde{\bar{\pmb{\Lambda}}}}(x')
         \hat{\Tilde{\pmb{\Lambda}}}(x) \Big\rangle
\,.
    \end{split}
\quad
%\nonumber
\end{equation}
Making use of Eqs.~(\ref{5.5}--\ref{5.6}) and~\eqref{5.9} we can write this 
as follows,
\begin{equation}
    \label{5.24}
    \begin{split}
        i\mathbf{S}_F(x;x') = \mathbf{X}^{\dagger}\cdot
         \big\langle \text{T}\big[\hat{\pmb{\Lambda}}^{(d)}(x)
          \hat{\bar{\pmb{\Lambda}}}^{(d)}(x')\big]\big\rangle\cdot \mathbf{Y}\\
        = \mathbf{X}^{\dagger}\cdot i\mathbf{S}_F^{(d)}(x;x')\cdot \mathbf{Y}
\,,
    \end{split}
\qquad
\end{equation}
where the propagator $i\mathbf{S}^{(d)}$ is in the mass diagonal basis and satisfies the following equation,
\begin{equation}
    \label{5.25}
    \left(i\gamma^{b}\partial_{b} - a\lvert \mathcal{M}^{(d)}\rvert\right)_{_{IP}}
    iS^{(d)}_{_{PJ}}(x;x') =i \delta_{IJ}\delta^{D}(x\!-\!x'),
\end{equation}
and $\lvert\mathcal{M}^{(d)}\rvert$ (see Eqs.~(\ref{D.3}--\ref{D.4}) in Appendix~E) 
is the real mass matrix obtained after diagonalization and is defined as,
\begin{equation}
    \label{5.26}
    \lvert\pmb{\mathcal{M}}^{(d)}\rvert = \begin{pmatrix}
    \lvert \textbf{m}^{(d)} \rvert & 0\\
    0 & \lvert \textbf{m}^{(d)} \rvert
    \end{pmatrix}
\,.\quad
\end{equation}
Thus we can see that the propagator in Eq. \eqref{5.24} can be rotated to remove the phases arising from $\mathbf{X}$ and $\mathbf{Y}$, therefore these phases are not observable (at tree level) 
and there is no CP-violation at tree level. 
Finally from the mode function solutions in Eq.~(\ref{5.18}--\ref{5.21}) we can find the propagator solution by following the same method that we used in section~\eqref{Section-3} to obtain,
\begin{equation}
    \label{5.27}
    i\mathbf{S}_F(x;x') = \mathbf{X}^{\dagger}\!\cdot\!\left\{a\left(i\gamma^{\mu}\nabla_{\mu} 
      \!+\! \lvert \pmb{\mathcal{M}}^{(d)}\rvert\right)\frac{H^{D-2}}{\sqrt{aa'}}
     \left[\frac{1\!+\!\gamma^{0}}{2} i\mathbf{S}^{+}(x;x')
      \!+\!\frac{1\!-\!\gamma^{0}}{2} i\Tilde{\mathbf{S}}^{+}(x;x')\right]\right\}\!\cdot\!\mathbf{Y}
,\qquad
\end{equation}
where $i\mathbf{S}^{+}(x;x')$ and $i\Tilde{\mathbf{S}}^{+}(x;x')$
denote a (diagonal) matrix generalization of solutions in Eq.~\eqref{3.12},
\begin{equation}
    \label{5.28}
    \begin{split}
        i\mathbf{S}^{+}(x;x') &= \frac{\Gamma\left(\frac{D}{2} \!+\! i\pmb{\zeta}^{(d)}\right)
        \Gamma\left(\frac{D-2}{2} \!-\!  i\pmb{\zeta}^{(d)}\right)}{(4\pi)^{D/2}
        \Gamma\big(\frac{D}{2}\big)}
        \times{}_{2}\pmb{F}_{1}\left(\frac{D}{2}\!+\!i\pmb{\zeta}^{(d)},
        \frac{D\!-\!2}{2}\!-\! i\pmb{\zeta}^{(d)};\frac{D}{2};1-\frac{y_{++}(x;x')}{4}\right)
\\
        i\Tilde{\mathbf{S}}^{+}(x;x') 
        &= \frac{\Gamma\left(\frac{D}{2}\!-\! i\pmb{\zeta}^{(d)}\right)
        \Gamma\left(\frac{D\!-\!2}{2} \!+\! i\pmb{\zeta}^{(d)}\right)}
        {(4\pi)^{D/2}\Gamma\big(\frac{D}{2}\big)}
        \times{}_{2}\pmb{F}_{1}\left(\frac{D}{2}\!-\! i\pmb{\zeta}^{(d)},
        \frac{D-2}{2}\!+\!i\pmb{\zeta}^{(d)};\frac{D}{2};1\!-\! \frac{y_{++}(x;x')}{4}\right)
\,,
    \end{split}
\end{equation}
where $\pmb{\zeta}^{(d)}$ is the diagonal matrix defined in Eq.~(\ref{5.22A}).

\noindent
\medskip

For completeness and for comparison, 
here we sketch how to construct the multiflavor Dirac propagator $i\mathbf{G}_F(x;x')$.
Firstly, one introduces unitary rotation matrices $\mathbf{R}$ and $\mathbf{Q}$,
which act on the propagator as, 
%(see Eqs.~ (\ref{E.16}--\ref{E.17}), and~(\ref{E.52}--\ref{E.53})),
%
\begin{equation}
    \label{5.31}
    \mathbf{R}\!\cdot\!i\mathbf{G}_F(x;x')\!\cdot\!\mathbf{Q} = i\mathbf{G}^{(d)}(x;x')
\,,\qquad
\end{equation}
where $ i\mathbf{G}^{(d)}(x;x')$ denotes the diagonal Dirac propagator
and the rotation matrices are given by,
\begin{equation}
    \label{5.32}
        \mathbf{R} = {\rm e}^{\frac{i}{2}\pmb{\theta}^{(d)}\gamma^{5}}\mathbf{V}
\,,\qquad
     \mathbf{Q} = \mathbf{U}^{\dagger}{\rm e}^{-\frac{i}{2}\pmb{\theta}^{(d)}\gamma^{5}}
\,,\qquad
\end{equation}
where
\begin{equation}
 {\rm e}^{\pm\frac{i}{2}\pmb{\theta}^{(d)}\gamma^{5}}
     = \cos\left(\frac{\pmb{\theta}^{(d)}}{2}\right)
     \pm i\gamma^{5}\sin\left(\frac{\pmb{\theta}^{(d)}}{2}\right)
\label{diagonal chiral matrices}
\end{equation}
are the diagonal chiral matrices with $n$ arbitrary phases, which can be used to rotate away
the phases from the diagonal elements of ${\mathcal{M}}^{(d)}$.
The propagator can then be written as, 
\begin{eqnarray}
        i\mathbf{G}(x;x') \!\!&=&\!\! \mathbf{V}^{\dagger}
               {\rm e}^{-\frac{i}{2}\pmb{\theta}^{(d)}\gamma^{5}}
        \left\{a\left(i\gamma^{\mu}\nabla_{\mu} 
        \!+\! \lvert \pmb{\mathcal{M}}^{(d)}\rvert\right)\frac{H^{D-2}}{\sqrt{aa'}}
        \left[\frac{1\!+\!\gamma^{0}}{2} i\pmb{S}^{+}(x;x')
         \!+\!\frac{1\!-\!\gamma^{0}}{2} i\Tilde{\pmb{S}}^{+}(x,x')\right]\right\}
\nonumber\\
       &\times&  {\rm e}^{\frac{i}{2}\pmb{\theta}^{(d)}\gamma^{5}}\mathbf{U}
\,,
    \label{5.33}
\\
\!\!&&\!\! \hskip -1.8cm
=\,\mathbf{V}^{\dagger}
        \left\{a\left(i\gamma^{\mu}\nabla_{\mu} 
        \!+\!  \pmb{\mathcal{M}}^{(d)}\right)\frac{H^{D-2}}{\sqrt{aa'}}
\left[\frac{1\!+\! {\rm e}^{i\pmb{\theta}^{(d)}\gamma^{5}}\gamma^{0}}{2}
            i\pmb{S}^{+}(x;x')
\!+\!\frac{1\!-\! {\rm e}^{i\pmb{\theta}^{(d)}\gamma^{5}}\gamma^{0}}{2}
        i\Tilde{\pmb{S}}^{+}(x,x')\right]\right\} \mathbf{U}
,
% \label{5.33b}
\nonumber
\end{eqnarray}
where $\pmb{\mathcal{M}}^{(d)}$ is a diagonal complex matrix, 
$i\pmb{S}^{+}(x;x')$ and $i\Tilde{\pmb{S}}^{+}(x;x')$ are defined in Eq.~(\ref{5.28})
and the unitary rotation matrices $\mathbf{U}$ and $ \mathbf{V}$ are given by,
\begin{equation}
    \label{5.34}
    \mathbf{U}  = \begin{pmatrix}
    \mathbf{U_{_{R}}} & 0 \\
    0 & \mathbf{U_{_{L}}}
    \end{pmatrix}
\,,\qquad
    \mathbf{V}^{\dagger}= \begin{pmatrix}
    \mathbf{U}^{\dagger}_{_{L}} & 0 \\
    0 & \mathbf{U}^{\dagger}_{_{R}}
    \end{pmatrix}
\,.\qquad
\end{equation}
One should keep in mind that $\mathbf{U}$ and $ \mathbf{V}$
are not general unitary matrices, but from each of them 
 $n$ common phases have been removed by factoring out the diagonal matrices 
${\rm e}^{\frac{i}{2}\pmb{\theta}^{(d)}\gamma^{5}}$. This means 
that we have used $2n^2$ parameters from the two unitary matrices to diagonalize
the Dirac mass matrix, for which $n^2 + n(n-1)=2n^2-n$ parameters are needed ($n^2$ 
do diagonalize the symmetric part of the mass matrix and $n(n-1)$ to diagonalize the 
antisymmetric part), such that $n$ parameters remain unused. 
 However, none of these 
$n$ parameters can be used to remove phases in the Yukawa mass matrix. 
To see this let us first look at the simplest case when $n=1$, the rotation
`matrices' have the form, $U_L={\rm e}^{i\phi}$ and $U_R={\rm e}^{i\psi}$, such that 
a complex mass $m=m_1+im_2=\rho {\rm e}^{i\mu}$ can be diagonalized
by $U_L m(U_R)^\dagger={\rm e}^{i\phi}\rho{\rm e}^{i\mu}{\rm e}^{-i\psi}=\rho$.
This fixes the phase difference, $\psi-\phi = \mu$, 
but leaves their sum $\phi+\psi$ unconstrained.
In the general case of $n$ Dirac flavors we have $m_{ij}=\rho_{ij}{\rm e}^{i\mu_{ij}}$. 
Let us for simplicity consider how the unitary matrices
of the form $(U_R)_{ij}={\rm e}^{i\phi_j}\delta_{ij}$  and 
$(U_L)_{ij}={\rm e}^{i\psi_j}\delta_{ij}$ (such that 
$(U_L^\dagger)_{ij}={\rm e}^{-i\psi_i}\delta_{ij}$) act,
\begin{eqnarray}
[(U_R)\cdot m\cdot U_L^\dagger]_{ij}
   \!\!&=&\!\! (U_R)_{ik}m_{kl}(U_L^\dagger)_{lj}
   =\delta_{ik}{\rm e}^{i\phi_k}\rho_{kl}{\rm e}^{i\mu_{kl}}{\rm e}^{-i\psi_l}\delta_{lj}
%\nonumber\\
  %  \!\!&=&\!\! 
  = {\rm e}^{i\phi_i}\rho_{ij}{\rm e}^{i\mu_{ij}}{\rm e}^{-i\psi_j}
\,.
\label{redundant parameters: diagonalization}
\end{eqnarray}
This then implies that, out of the $2n$ phases $\{\phi_i,\psi_i\}\, (i=1,2,\cdots,n)$,
one can only use $n$ of the differences $\phi_i-\psi_j$ to remove some of the
phases in $\mu_{ij}$. 
However, the $n$ remaining 
linearly independent combinations $\phi_i+\psi_j$ cannot be used to remove any phases 
as they appear with a `wrong sign' in~(\ref{redundant parameters: diagonalization}).
For the same reason, none of the remaining $n$ phases $\phi_i+\psi_j$ can be used to 
remove any of the phases in the Yukawa matrix $y_{ij}$
({\it cf.} Eq.~(\ref{yukawa matrix: def})).
This means that 
the group that diagonalizes a general Dirac mass matrix is 
not direct product of the two unitary groups,
${\cal U}_L(n)\otimes {\cal U}_R(n)$, but instead it is the quotient group,
$[{\cal U}_L(n)\otimes {\cal U}_R(n)]/{\cal U}(1)^n$.
This consideration also shows that, even though after the diagonalization 
of the mass matrix the unitary matrices $U_L$ and $U_R$ possess 
$n$ redundant parameters, none of them can be used to remove any of the phases from 
the Yukawa matrix. 
With this remark we conclude our analysis of the Dirac fermion propagator for mixing flavors,
which shows that there are CP violating effects at the tree level, as was expected.

\bigskip

To understand the question of CP violation a bit better, 
let us summarize the principal differences between the Majorana
and Dirac fermions. The mass matrix and Yukawa coupling matrix for multiflavor 
Majorana fermions are complex symmetric matrices which consist of $n(n-1)$ 
off-diagonal real parameters and 
$n$ diagonal phases (totalling $n^2$ parameters), which can be all removed by 
the Tagaki diagonalization procedure by making use of a unitary matrix 
which contains $n^2$ real parameters.
Since this completely fixes the unitary matrix, diagonalization of the mass matrix
leaves the Majorana Yukawa matrix completely general. Such a symmetric complex matrix has 
$n(n+1)/2$ phases and $n(n+1)/2$ real parameters, meaning that 
Majorana fermions  can harbor $n(n+1)/2$ CP-violating phases in the Yukawa matrix.
This is to be contrasted with the Dirac fermions, for which both the mass matrix and 
Yukawa matrix are general complex matrices with $2n^2$ real parameters in total. 
The diagonalization requires 
two independent unitary matrices, each of them having $n^2$ real parameters, such 
that diagonalization requires $2n^2-n$ real parameters (after diagonalization $n$ real 
eigenvalues remain), leaving $n$ parameters unfixed.
As we have shown above, all of these $n$ parameters are redundant, meaning 
that they cannot be used to remove any of the CP phases in the Yukawa matrix.
The general Yukawa matrix for Dirac fermions contains $2n^2$ parameters,
which can be decomposed into a symmetric matrix (with $n(n+1)$ real parameters,
$n(n+1)/2$ are phases) and an antisymmetric matrix (with $n(n-1)$ real parameters,
$n(n-1)/2$ are phases), such that in general Dirac fermions harbor at most
 $n(n+1)/2+n(n-1)/2=n^2$ CP violating phases. When this is compared with Majorana
  fermions, Dirac fermions can harbor $n^2-n(n+1)/2=n(n-1)/2$ more CP violating
  phases than Majorana fermions.

%%%%%%%%%%%%%%%%%%%%%%%%%%%%%%
%%%    C O N C L U S I O N   A N D   D I S C U S S I O N   %%%
%%%%%%%%%%%%%%%%%%%%%%%%%%%%%%

\section{Conclusion and Discussion}
\label{Conclusion and Discussion}

In this paper we construct the Majorana propagator in de Sitter space for a complex
Majorana mass~(\ref{3.12}--\ref{3.11}),
as well as for more general multiflavor Majorana 
fields~(\ref{5.27}--\ref{5.28}).~\footnote{Our results differ from earlier 
works~\cite{Cotaescu:2018zrq,Cotaescu:2018afn}, where 
the Majorana propagator was constructed by imposing the left chiral projector 
$P_L=(1-\gamma^5)/2$ on the Dirac propagator in de Sitter.
While this projection correctly reduces the number of degrees of freedom,
it is inconsistent with the Majorana condition, and therefore, in our opinion, questionable.}
Our work is relevant for the dynamics of neutrinos in the early Universe setting, 
and more generally for the dynamics of Majorana particles that may be 
involved {\it e.g.} in leptogenesis scenarions, which constitute a popular 
explanation for the observed matter-antimatter asymmetry of the Universe.
In spite of various subtleties  involved in construction 
of the Majorana propagator~\footnote{These subtleties include an
off-shell imposition of the Majorana condition,
a careful account of canonical quantization, a splitting of the momentum space 
mode functions into two hemispheres, {\it etc.}}, 
the final result~(\ref{3.12}--\ref{3.11}) is identical to the Dirac propagator.
This can be understood as follows. Even though at the operator level 
there is a clear distinction between the Dirac fermions (for which the positive and negative 
frequency poles fluctuate independently) and the Majorana fermions (for which the 
same operator is responsible for fluctuations at the positive and negative 
frequency poles), their statistical properties are the same because the (gravitational)
 coupling of fermions
to de Sitter space does not violate CP symmetry. In other words, differences between
Majorana and Dirac fermions would arise in CP-violating backgrounds.
Such CP-violating backgrounds can occur, for example, during first order phase transitions
in the early Universe, and which may be suitable for dynamical baryogenesis.
One example of such a CP-violating background was considered 
in Ref.~\cite{Prokopec:2013ax}, in which the dynamics of Dirac fermions 
in presence of such a CP-violating background was studied, and in which indeed 
Dirac fermions exhibit CP-violating effects. Our generalization to 
the case of multiflavor Majorana fields~(\ref{5.27}--\ref{5.28}) was done 
in the prescience of quantum loop studies, in which the Majorana phases are expected 
to show CP-violating effects. For comparison, we also quote the mutiflavor Dirac propagator,
and emphasise that mixing Dirac fermions 
contain more CP violating phases than Majorana fermions,
which could be a way to distinguish whether the nature of fermions is Dirac or Majorana.
One of these phases is the chiral phase, $\phi = {\rm Arctan}[m_2/m_1]$,
 of the complex mass term $m=m_1+im_2$, which is present already at the single fermion
level. Curiously, this phase modifies the positive and negative frequency projectors,
$P_\pm = (1\pm\gamma^0)/2$, in the propagator~(\ref{3.11}) 
 to give them a chiral nature, 
$P^5_\pm = \big(1\pm{\rm e}^{i\gamma^5\phi}\gamma^0\big)/2$.
Understanding the full significance of this observation requires quantum loop studies, 
and it is thus beyond the scope of this work.

Even though the Dirac and Majorana vacuum propagators in de Sitter are identical,
one ought to be careful when using them in loop studies. To illustrate this, 
in section~\ref{One-Loop Effective Action}
we compute the one-loop effective action
$\Gamma^{(1)}_{\text{Maj}}$  for Majorana particles~(\ref{renormalized action}),
 and show that it differs from that of Dirac fermions
(calculated earlier in {\it e.g.} Refs.~\cite{Candelas:1975du,Koksma:2009tc})
 by a factor 1/2, 
\begin{equation}
    \label{6.1}
    \Gamma^{(1)}_{\text{Maj}} = \frac{1}{2} \Gamma^{(1)}_{\text{Dirac}}
\,,
\end{equation}
which was to be expected since the Majorana fermions carry 1/2 of the degrees of freedom
when compared with the Dirac fermions. 

%Understanding further differences
%between the Dirac and Majorana fermions requires studying of other loop processes.

\vspace{5mm}

%%%%%%%%%%%%%%%%%%%%%%%
%%%    A C K N O W L E D G E M E N T S   %%%
%%%%%%%%%%%%%%%%%%%%%%%

\noindent
\section*{Acknowledgments.} 
The authors are grateful to Tanja Hinderer and Elisa Chisari for useful comments.
This work was partially supported by the D-ITP consortium, a
program of the Netherlands Organization for Scientific Research (NWO) that
is funded by the Dutch Ministry of Education, Culture and Science (OCW).

%%%%%%%%%%%%%%%%%%
%%%      A P P E N D I C E S     %%%
%%%%%%%%%%%%%%%%%%

%\begin{appendices}

\section{Appendices}
\label{Appendices}

\section*{Appendix A: Dirac's matrices in chiral representation}
\label{Appendix-C}

In this paper we use chiral representation for Dirac's gamma matrices, which in flat (tangent) space
and in $D=4$ are of the form,
\begin{equation}
    \label{gamma:A.1}
    \gamma^a =\begin{pmatrix}
    0 & \sigma^a \\
    \bar \sigma^a & 0\\
    \end{pmatrix} 
\,,\qquad \sigma^a = \left(\mathbb{1},\sigma^i\right)
\,,\qquad \bar\sigma^a = \left(\mathbb{1},-\sigma^i\right)
\,,
\qquad
(a=0,1,2,3)
\,,\qquad
\end{equation}
where $\sigma^i$ are Pauli matrices. They obey the anti-commutation relation,
$\{\gamma^a,\gamma^b\}=2\eta^{ab}$ and $\gamma^5 = {\rm diag}(-\mathbb{1},\mathbb{1})$.
The following commutation properly is useful for Majorana fermions, 
\begin{equation}
   \label{gamma:A.2}
 \gamma^0 (-i\gamma^2)\big[\gamma^a\big]^T 
  = - \gamma^a\gamma^0 (-i\gamma^2)
\,.\qquad
\end{equation}
Charge conjugation operator is given by the following,
\begin{equation}
     \label{gamma:A.3}
     \begin{split}
         \mathcal{C} = - i\gamma^{2} = \begin{pmatrix}
     0 & -\epsilon\\
     \epsilon & 0
     \end{pmatrix} \,,\qquad \epsilon = i\sigma^{2}  = \begin{pmatrix}
     0 & 1 \\ 
     -1 & 0
     \end{pmatrix} \,,\qquad \epsilon^{2} = -\mathbf{1}_{2 \times 2}\\
     \mathcal{C}^{\dagger} = \mathcal{C}\qquad\qquad\qquad\qquad\qquad\qquad
     \end{split}
\end{equation}
where $\epsilon$ is an antisymmetric matrix and the following property satisfied by the charge conjugation operator comes in handy
\begin{equation}
    \label{gamma:A.4}
    (-i\gamma^{2})(-i\gamma^{2})^{\dagger} = (-i\gamma^{2})(-i\gamma^{2}) 
          = \mathbf{1}_{4\times 4}.
\end{equation}
In what follows we introduce some transformations of projection operators, 
%$P_{\pm}$ and $P^{5}_{\pm}$ which are given by
%
\begin{equation}
    \label{gamma:A.5}
%    \begin{split}
        P_{\pm} = \frac{1\pm \gamma^{0}}{2}
\,,\qquad
        P^{5}_{\pm} = \frac{1\pm e^{i\theta\gamma^{5}} \gamma^{0}}{2}.
%    \end{split}
\end{equation}
These operators transform under the charge conjugation as follows,
\begin{equation}
    \label{gamma:A.6}
        \mathcal{C} P_{\pm} \mathcal{C}^{\dagger} = P_{\mp}
\,,\qquad
        \mathcal{C} P^{5}_{\pm} \mathcal{C}^{\dagger} = \left(P^{5}_{\mp}\right)^{\dagger}
\,.\quad
\end{equation}

\section*{Appendix B: Mode decomposition for Majorana fields}
\label{Appendix-C}

The rescaled classical Dirac field $\Tilde{\Psi}_{\alpha,D}(x)$ can be represented in the spatial momentum space as folows,
\begin{equation}
    \label{C.2}
    \hat{\Tilde{\Psi}}_{\alpha,D}(x) = \begin{pmatrix}
    \hat{\Tilde{\chi}}_{_{L}}(x)\\
    \hat{\Tilde{\chi}}_{_{R}}(x)
    \end{pmatrix} = \int \frac{d^{D-1}k}{(2\pi)^{D-1}}{\rm e}^{i\vec{k}\cdot\vec{x}}C_{h}(\eta, \vec{k})
\,,
\end{equation}
where $\hat{\Tilde{\chi}}_{_{L}}(x)$ and $\hat{\Tilde{\chi}}_{_{R}}(x)$ are the rescaled 
chiral 2-spinors and $C_{h}(\eta, \vec{k})$ is the momentum and helicity eigenspinor defined by,
\begin{equation}
    \label{C.3}
    C_{h}(\eta, \vec{k}) = \begin{pmatrix}
    L_{h}(\eta, \vec{k})\xi_{h}(\vec{k})\\
    R_{h}(\eta, \vec{k})\xi_{h}(\vec{k})
    \end{pmatrix},
\end{equation}
where  $L_{h}(\eta, \vec{k})$ and  $R_{h}(\eta, \vec{k})$ 
are complex chiral mode functions $\xi_{h}(\vec{k})$
are the helicity eigenspinors, whose detailed properties are discussed in Appendix~C. 
The decomposition~(\ref{C.2}) is particularly convenient in spatially homogeneous spaces such 
as de Sitter space, as ${\rm e}^{i\vec{k}\cdot\vec{x}}$ are eigenstates of the spatial 
derivative operator $\partial_i$ with the eigenvalue $ik_i$.
It is convenient (and customary) to recast the rescaled quantum Dirac field 
$\hat{\Tilde{\Psi}}_{\alpha,D}(x)$ 
in terms of the mode operators decomposed into distinct creation 
($\hat{d}^{\,\dagger}_{h}(\vec{k})$)
 and annihilation ($\hat{b}_{h}(\vec{k})$) operators,
\begin{equation}
    \label{C.4}
    \hat{\Tilde{\Psi}}_{\alpha,D}(x) = \sum_{h=\pm}\int\frac{d^{D-1}k}{(2\pi)^{D-1}}e^{i\vec{k}\cdot\vec{x}}\left[C_{h,\alpha}(\eta, \vec{k})\hat{b}_{h}(\vec{k})
     + D_{h,\alpha}(\eta, -\vec{k})\hat{d}^{\,\dagger}_{h}(-\vec{k})\right],
\end{equation}
where $D_{h,\alpha}(\eta, -\vec{k}) = -i\gamma^{2}C^{*}_{h,\alpha}(\eta, -\vec{k})$. In contrast, the rescaled Majorana fermions can be described by imposing the Majorana condition on Eq. \eqref{C.2}, 
\begin{equation}
    \label{C.5}
    \hat{\Tilde{\Psi}}_{\alpha,D}(x) = -i\gamma^{2}\hat{\Tilde{\Psi}}^{*}_{\alpha,D}(x)
    \,.
\end{equation}
From this condition we find the following relation between the two spinors that form the Dirac spinor in Eq.~\eqref{C.3}
\begin{equation}
    \label{C.6}
    \hat{\Tilde{\chi}}_{_{R}}(x) = \epsilon\hat{\Tilde{\chi}}_{_{L}}^{*}(x).
\end{equation}
After imposing the condition in Eq. \eqref{C.5} and the subsequent relation in Eq. \eqref{C.6} in Eq. \eqref{C.2} we have for the rescaled Majorana field $\hat{\Tilde{\Psi}}_{M,\alpha}(x)$,
\begin{equation}
    \label{C.7}
    \hat{\Tilde{\Psi}}_{D,\alpha}(x)\;\stackrel{\rm Maj.\; condition}{\longrightarrow}\;\hat{\Tilde{\Psi}}_{M,\alpha}(x) = \begin{pmatrix}
    \hat{\rho}(x)\\
    \epsilon\hat{\rho}^{*}(x)
    \end{pmatrix}
    \,,
\end{equation}
Thus the classical momentum decomposition for $\hat{\Tilde{\Psi}}_{M,\alpha}(x)$ is defined by, 
\begin{equation}
    \label{C.10}
    \hat{\Tilde{\Psi}}_{M,\alpha}(x) 
    = \int\frac{d^{D-1}k}{(2\pi)^{D-1}}e^{i\vec{k}\cdot\vec{x}}A_{h}(\eta, \vec{k})
\,,
\end{equation}
where the momentum eigenspinor $A_{h}(\eta, \vec{k})$, which gives the positive energy solutions, is defined as,
\begin{equation}
    \label{C.11}
    A_{h}(\eta, \vec{k}) = \begin{pmatrix}
    L_{h}(\eta, \vec{k}) \xi_{h}(\vec{k})\\
    L^{*}_{h}(\eta, -\vec{k}) \epsilon\xi^{*}_{h}(-\vec{k})
    \end{pmatrix}
\,.
\end{equation}
The quantized rescaled Majorana fermion field is then decomposed as,
\begin{equation}
    \label{C.12}
    \hat{\Tilde{\Psi}}_{M,\alpha}(x) = \sum_{h=\pm}\int \frac{d^{D-1}k}{(2\pi)^{D-1}}
    \left[{\rm e}^{i\vec{k}\cdot\vec{x}}A_{h}(\eta, \vec{k})\hat b_{h}(\vec{k})
     + {\rm e}^{-i\vec{k}\cdot\vec{x}}B_{h}(\eta, \vec{k})\hat b^{\dagger}_{h}(\vec{k})\right],
\end{equation}
with $\hat b_{h}(\vec{k})$ and $\hat b^{\dagger}_{h}(\vec{k})$ are the identical 
annihilation and creation operators. 
The momentum eigenspinor $B_{h}(\eta, \vec{k})$ obeys the Majorana condition,
\begin{equation}
    \label{C.13}
    B_{h}(\eta, \vec{k}) = -i\gamma^{2}A^{*}_{h}(\eta, \vec{k}) = A_{h}(\eta, -\vec{k})
\,.
\end{equation}
We can now use this result to write Eq. \eqref{C.12} as,
\begin{equation}
    \label{C.14}
    \begin{split}
        \hat{\Tilde{\Psi}}_{M,\alpha}(x) &= \sum_{h=\pm}\int \frac{d^{D-1}k}{(2\pi)^{D-1}}\left[e^{i\vec{k}\cdot\vec{x}}A_{h}(\eta, \vec{k})\hat b_{h}(\vec{k})
         + e^{-i\vec{k}\cdot\vec{x}}A_{h}(\eta, -\vec{k})\hat b^{\dagger}_{h}(\vec{k})\right]\\
        &= \sum_{h=\pm}\int \frac{d^{D-1}k}{(2\pi)^{D-1}}e^{i\vec{k}\cdot\vec{x}}\left[\hat b_{h}(\vec{k}) + \hat b^{\dagger}_{h}(-\vec{k})\right]A_{h}(\eta, \vec{k})
\,,
    \end{split}
\end{equation}
which is the form used in the main text in Eqs.~(\ref{2.18}--\ref{2.18b}).

\section*{Appendix C: Properties of helicity 2-eigenspinor}
\label{Properties of Helicity 2-spinor}
\label{Appendix-A}

Here we elucidate the properties of the helicity 2-eigenspinor in $D=4$ dimensions for the purposes of keeping the calculations simple. This can be extended to 
D dimensions as was done in Ref.~\cite{Koksma:2009tc}. 
We have seen the definition of helicity 2-eigenspinor 
in Eq.~\eqref{2.23}. However $\epsilon\xi^{*}_h(-\vec{k})$ can be written as follows
\begin{equation}
    \label{A.1}
    \begin{split}
        \epsilon\xi^{*}_{h}(-\vec{k}) = \frac{1}{\sqrt{2\left(1 + h\hat{k}_{z}\right)}}\begin{pmatrix}
    0 & 1\\
    -1 & 0
    \end{pmatrix}\begin{pmatrix}
    -h\left(\hat{k}_{x} + i\hat{k}_{y}\right)\\
    1 + h\hat{k}_{z}
    \end{pmatrix}\\
     \implies \epsilon\xi^{*}_{h}(-\vec{k}) = \frac{1}{\sqrt{2\left(1 + h\hat{k}_{z}\right)}} \begin{pmatrix}
     1 + h\hat{k}_{z}\\
     h\left(\hat{k}_{x} + i\hat{k}_{y}\right)
     \end{pmatrix}.
    \end{split}
\end{equation}
After massaging the terms in the matrix in Eq.\eqref{A.1}, we can write it as follows
\begin{equation}
    \label{A.2}
    \epsilon\xi^{*}_{h}(-\vec{k}) = h\sqrt{\frac{\hat{k}_{x} + i\hat{k}_{y}}{\hat{k}_{x} - i\hat{k}_{y}}}\xi_{h}(\vec{k}) 
\end{equation}
 Then we can write the factor $h\sqrt{\frac{\hat{k}_{x} + i\hat{k}_{y}}{\hat{k}_{x} - i\hat{k}_{y}}}$ as a phase,
\begin{equation}
    \label{A.4}
    \sqrt{\frac{\hat{k}_{x} + i\hat{k}_{y}}{\hat{k}_{x} - i\hat{k}_{y}}} = \sqrt{\frac{\lvert \Tilde{k} \rvert e^{i\Tilde{\theta}}}{\lvert \Tilde{k} \rvert e^{-i\Tilde{\theta}}}} = e^{i{\theta}(\hat{k}_{x},\hat{k}_{y})},
\end{equation}
where we have defined, 
$\lvert \Tilde{k} \rvert = \sqrt{\hat{k}^{2}_{x} + \hat{k}^{2}_{y}}$ 
and $\theta(\hat{k}_{x},\hat{k}_{y}) = \tan^{-1}\left(\frac{\hat{k}_{y}}{\hat{k}_{x}}\right)$. 
Now we can see that
\begin{equation}
    \label{A.5}
    \epsilon\xi^{*}_{h}(-\vec{k}) = he^{i\theta(\hat{k}_{x},\hat{k}_{y})}\ \xi(\vec{k}) 
\end{equation}
\begin{equation}
\label{A.6}
    (\epsilon\xi^{*}_{h}(-\vec{k}))^{\dagger} =he^{-i\theta(\hat{k}_{x},\hat{k}_{y})} \xi^{\dagger}(\vec{k}),
\end{equation}
where we have made use of Eq.\eqref{A.4}. We also find the following property useful
\begin{equation}
\label{A.7}
    \sum_{h=\pm}\xi_{h,a}^{\dagger}(\vec{k})\otimes\xi_{h,b}(\vec{k}) = \delta_{ab}
\,, 
\end{equation}
which can be proved for the helicity 2-eigenspinor ($a,b\in\{1,2\}$) as follows,
\begin{equation}
\label{A.12}
\begin{split}
\sum_{h=\pm}\xi_{h,a}^{\dagger}(\vec{k})\otimes\xi_{h,b}(\vec{k}) 
&= \sum_{h=\pm} \frac{1}{2\left(1-h\hat{k}_{z}\right)}\begin{pmatrix}
h\left(\hat{k}_{x} + i\hat{k}_{y}\right) & 1 - h\hat{k}_{z}
\end{pmatrix}\otimes
\begin{pmatrix}
h\left(\hat{k}_{x} - i\hat{k}_{y}\right)\\
\left(1 - h\hat{k}_{z}\right)
\end{pmatrix} 
\\
&= \sum_{h=\pm}\begin{pmatrix}
\frac{1 + h\hat{k}_{z}}{2}  &  \frac{h\left(\hat{k}_{x} - i\hat{k}_{y}\right)}{2}\\
\frac{h\left(\hat{k}_{x} + i\hat{k}_{y}\right)}{2}  &  \frac{1 - h\hat{k}_{z}}{2}
\end{pmatrix}
%\hspace{6cm}\\
= \begin{pmatrix}
1 & 0 \\
0 & 1
\end{pmatrix} = \delta_{ab}
\,.\hspace{8.5cm}
\end{split}
\end{equation}
We also make note of the following property which is going to be crucial when we discuss the equations of motion in section II~B~3. $he^{i\theta(\hat{k}_{x},\hat{k}_{y})}\xi_{h}(\vec{k})$ under $\vec{k}\rightarrow -\vec{k}$ transforms as follows,
\begin{equation}
    \label{A.10}
    he^{i\theta\left(\hat{k}_{x},\hat{k}_{y}\right)}\xi_{h}(\vec{k})
    \; \xrightarrow{\vec{k}\rightarrow -\vec{k}}\;
     -he^{i\theta\left(\hat{k}_{x},\hat{k}_{y}\right)}\xi_{h}(-\vec{k})
\,.
\end{equation}
This can be seen either by calculating $\epsilon\xi^{*}_{h}(\vec{k})$,
 which yields $ -he^{i\theta\left(\hat{k}_{x},\hat{k}_{y}\right)}\xi_{h}(-\vec{k})$,
or we can understand this from a  topological viewpoint where under 
$\vec{k}\rightarrow -\vec{k}$ we have the following transformation of the phase 
$\theta(\hat{k}_{x},\hat{k}_{y})$,
\begin{equation}
\label{A.11}
\theta(\hat{k}_{x},\hat{k}_{y})\; \xrightarrow{\vec{k}\rightarrow -\vec{k}}\;\theta(\hat{k}_{x}
\,,\hat{k}_{y}) + \pi
\,.
\end{equation}

\section*{Appendix D: Properties of Hankel Functions}
\label{Properties of Hankel Functions}
\label{Appendix-B}

For convenience here we list some of the properties of Hankel functions used in this work.
Firstly we have, 
\begin{equation}
\label{B.1}
    H^{(1)}_{-\nu}(z) = e^{i\pi\nu}H^{(1)}_{\nu}(z)
\end{equation}
\begin{equation}
\label{B.2}
    H^{(2)}_{-\nu}(z) = e^{-i\pi\nu}H^{(2)}_{\nu}(z)
\end{equation}
\begin{equation}
\label{B.3}
   \{H^{(1)}_{\nu}(z)\}^{*} = H^{(2)}_{\nu^{*}}(z^{*}) 
\end{equation}
\begin{equation}
\label{B.4}
    H_{\nu}^{(1)}(e^{i\pi}z) = -H^{(2)}_{-\nu}(z) = -e^{-i\pi\nu}H^{(2)}_{\nu}(z)
\,,
\end{equation}
\begin{equation}
\label{B.5}
    H_{\nu}^{(2)}(e^{-i\pi}z) = -H^{(1)}_{-\nu}(z) = -e^{i\pi\nu}H^{(1)}_{\nu}(z)
\,.
\end{equation}
The Wronskian and the recurrence relation are given by,
\begin{equation}
    W[H^{(1)}_{\nu},H^{(2)}_{\nu}] = -\frac{4i}{\pi z} \label{B.6}
\end{equation}
\begin{equation}
    H^{(i)}_{\nu - 1}(z) = \frac{d}{dz}H^{(i)}_{\nu}(z) + \frac{\nu}{z}H^{(i)}_{\nu}(z).\label{B.7}
\end{equation}
The Hankel functions are related to MacDonald functions $K_{\nu}(z)$ through 
the following identities,
\begin{equation}
    \label{B.8}
    H^{(1)}_{\nu}(z) = -\frac{2i}{\pi}e^{-\frac{i\pi\nu}{2}}K_{\nu}(-iz)
\end{equation}
\begin{equation}
    \label{B.9}
    H^{(2)}_{\nu}(z) = \frac{2i}{\pi}e^{\frac{i\pi\nu}{2}}K_{\nu}(iz).
\end{equation}
From Eqs.~(\ref{B.1}--\ref{B.7}) we can also write the following properties,
\begin{equation}
    \label{B.10}
    \begin{split}
        H^{(1)}_{\nu_{+}}(-\| \textbf{k} \|\eta) = - \frac{e^{i\pi\nu_{-}}}{\| \textbf{k} \|} \left[\partial_{\eta} + \frac{\nu_{-}}{\eta}\right]H^{(1)}_{\nu_{-}}(-\| \textbf{k} \|\eta)\\
         H^{(1)}_{\nu_{-}}(-\| \textbf{k} \|\eta) = - \frac{e^{i\pi\nu_{+}}}{\| \textbf{k} \|} \left[\partial_{\eta} + \frac{\nu_{+}}{\eta}\right]H^{(1)}_{\nu_{+}}(-\| \textbf{k} \|\eta)
    \end{split}
\end{equation}
\begin{equation}
    \label{B.11}
    \begin{split}
        H^{(2)}_{\nu_{+}}(-\| \textbf{k} \|\eta) = - \frac{e^{-i\pi\nu_{-}}}{\| \textbf{k} \|} \left[\partial_{\eta} + \frac{\nu_{-}}{\eta}\right]H^{(2)}_{\nu_{-}}(-\| \textbf{k} \|\eta)\\
         H^{(1)}_{\nu_{-}}(-\lvert \textbf{k}\rvert\eta) = - \frac{e^{-i\pi\nu_{+}}}{\| \textbf{k} \|} \left[\partial_{\eta} + \frac{\nu_{+}}{\eta}\right]H^{(2)}_{\nu_{+}}(-\| \textbf{k} \|\eta).
    \end{split}
\end{equation}

\section*{Appendix E: Diagonalization of the Majorana mass matrix}
\label{Diagonalization of Majorana mass matrix}
\label{Appendix-D}

We consider the mass matrix for the multiflavor Majorana mass from Eq.~\eqref{5.1} to be Hermitian and the Dirac equation in Eq. \eqref{5.4} can be written as,
\begin{equation}
    \label{D.1}
    \left(i\gamma^{b}\partial_{b}\pmb{1}
     - \mathbf{Y}\cdot\pmb{\mathcal{M}}\cdot\mathbf{X}^{\dagger}\right)
     \cdot\mathbf{\Lambda}^{(d)}(x) = 0
\,.
\end{equation}
Consider the term $\mathbf{Y}\cdot\pmb{\mathcal{M}}\cdot\mathbf{X}^{\dagger}$
(see Eq.~(\ref{mass matrix: def})):
\begin{equation}
    \label{D.2}
    \mathbf{Y}\!\cdot\!\pmb{\mathcal{M}}\!\cdot\!\mathbf{X}^{\dagger} 
    = \begin{pmatrix}
    \mathbf{U}^{*}_{_{L}} & 0\\
    0 & \mathbf{U}_{_{L}}
    \end{pmatrix}\!\cdot\!\begin{pmatrix}
    \mathbf{m}^{\dagger} & 0\\
    0 & \mathbf{m}
    \end{pmatrix}\!\cdot\! \begin{pmatrix}
    \mathbf{U}^{\dagger}_{_{L}} & 0\\
    0 & \mathbf{U}^{T}_{_{L}}
    \end{pmatrix}= \begin{pmatrix}
    \mathbf{U}^{*}_{_{L}}\!\cdot\!\mathbf{m}^{\dagger}\!\cdot\!\mathbf{U}^{\dagger}_{_{L}} & 0
    \\
    0 & \mathbf{U}_{_{L}}\!\cdot\!\mathbf{m}\!\cdot\!\mathbf{U}^{T}_{_{L}}
    \end{pmatrix},
\end{equation}
then we can use Takagi diagonalization~\cite{Hahn:2006hr,ChoiHaber} which says that for every 
symmetric complex mass matrix $\mathbf{m}$, there exists a unitary matrix 
$\mathbf{U}_{_{L}}$ such that,
\begin{equation}
    \label{D.3}
    \mathbf{U}_{_{L}}\!\cdot\!\mathbf{m}\!\cdot\!\mathbf{U}^{T}_{_{L}} 
    = \lvert \textbf{m}^{(d)} \rvert 
    = \text{diag}\left[m^{(d)}_{11},m^{(d)}_{22},\cdots,m^{(d)}_{nn}\right],
\end{equation}
where $\lvert\textbf{m}^{(d)}\rvert$ is real and non-negative as given from the Takagi theorem
(a special care must be taken when the mass matrix is degenerate~\cite{ChoiHaber}). 
And subsequently taking the hermitian conjugate of Eq.~\eqref{D.3} gives,
\begin{equation}
    \label{D.4}
    \mathbf{U}^{*}_{_{L}}\!\cdot\!\mathbf{m}^\dagger\!\cdot\!\mathbf{U}^{\dagger}_{_{L}} 
    = \lvert \textbf{m}^{(d)\dagger} \rvert = \lvert \textbf{m}^{(d)} \rvert,
\end{equation}
therefore, we can write Eq. \eqref{D.2} as follows
\begin{equation}
    \label{D.5}
    \mathbf{Y}\!\cdot\!\mathcal{M}\!\cdot\!\mathbf{X}^{\dagger}
    = \lvert \pmb{\mathcal{M}}^{(d)}\rvert = \begin{pmatrix}
    \lvert\mathbf{m}^{(d)}\rvert & 0\\
    0 & \lvert\mathbf{m}^{(d)}\rvert
    \end{pmatrix},
\end{equation}
and subsequently Eq.~\eqref{D.1} can be written as,
\begin{equation}
    \label{D.6}
    \left(\pmb{1}i\gamma^{b}\partial_{b} - \lvert \pmb{\mathcal{M}}^{(d)}\rvert\right)
    \!\cdot\!\mathbf{\Lambda}^{(d)}(x) = 0
\,.\qquad
\end{equation}
%

%\end{appendices}

 %%%%%%%%%%%%%%%%%%%%
%%%%   B I B L I O G R A P H Y    %%%%
 %%%%%%%%%%%%%%%%%%%%

\end{document}